\documentclass[12pt]{article}
\usepackage{fullpage, amsthm, amsmath}
\usepackage{mathtools}
\usepackage{graphicx, psfrag, amsfonts, bm, epsfig}
\usepackage{natbib}
\usepackage{subfigure}
\usepackage{hyperref}

\usepackage{graphicx, psfrag, amsfonts, setspace,lscape,longtable}
\usepackage{amssymb}
\usepackage{mathrsfs}
\usepackage{url}
\usepackage{relsize}
\usepackage{algorithm}

\usepackage{xr}
\newcommand*{\rom}[1]{\expandafter\@slowromancap\romannumeral #1@}

\newtheorem{lemma}{Lemma}

\newtheorem{remark}{Remark}

\newcommand{\is}{\itemsep=0pt}

\newcommand{\bd}[1]{\begin{description}[#1]\is}
\newcommand{\ed}{\end{description}}

\newcommand{\bi}{\begin{itemize}\is}
\newcommand{\ei}{\end{itemize}}

\newcommand{\be}{\begin{enumerate}\is}
\newcommand{\ee}{\end{enumerate}}

\newcommand{\beginsupplement}{
\setcounter{table}{0}
\renewcommand{\thetable}{S\arabic{table}}
\setcounter{figure}{0}
\renewcommand{\thefigure}{S\arabic{figure}}
\setcounter{equation}{0}
\renewcommand{\theequation}{S\arabic{equation}}
\setcounter{lemma}{0}
\renewcommand{\thelemma}{S\arabic{lemma}}
\setcounter{theorem}{0}
\renewcommand{\thetheorem}{S\arabic{theorem}}
\setcounter{section}{0}
\renewcommand{\thesection}{S\arabic{section}}
}

\DeclareMathOperator{\E}{E}
\DeclareMathOperator{\Var}{Var}

\DeclareMathOperator{\Length}{length}

\newcommand{\GaSP}{\text{GaSP}}
\newcommand{\N}{\text{N}}
\newcommand{\MN}{\text{MN}}

\def\beginmat{ \left( \begin{array} }
\def\endmat{ \end{array} \right) }

\usepackage[dvipsnames,usenames]{color}

\usepackage[title]{appendix}
\usepackage{etoolbox}
\patchcmd{\appendices}{\quad}{. }{}{}

\usepackage{tikz}
\usetikzlibrary{shapes,backgrounds}

\begin{document}
\doublespacing

\begin{titlepage}
\clearpage

\title{Fast Nonseparable Gaussian Stochastic Process with Application to Methylation Level Interpolation}

\date{}
\author{Mengyang Gu$^*$ and Yanxun Xu$^{**}$\\
           \small $^{*}$ Department of Statistics and Applied Probability, University of California, Santa Barbara, CA \\
        \small    $^{**}$ Department of Applied Mathematics and Statistics, Johns Hopkins University, Baltimore, MD
           }
\maketitle
\thispagestyle{empty}

\begin{abstract} 
 Gaussian stochastic process (GaSP) has been widely used as a prior over functions due to its flexibility and tractability in modeling.  However, the computational cost in evaluating the likelihood is $O(n^3)$, where $n$ is the number of observed points in the process, as it requires to invert the covariance matrix.  This bottleneck prevents GaSP being widely used in large-scale data.  We propose a general class of nonseparable GaSP models for multiple  functional observations  with a fast and exact algorithm, 
 in which the computation is linear ($O(n)$) and exact, requiring no  approximation to compute the likelihood. We show that the  commonly used  linear regression and separable models are special cases of the proposed nonseparable GaSP model.  Through the study of an epigenetic application, the proposed nonseparable GaSP model can accurately predict the genome-wide DNA methylation levels and compares favorably to alternative methods, such as linear regression, random forest and localized Kriging method. The  algorithm for fast computation is implemented in the ${\tt FastGaSP}$ R package on CRAN. 
 
\noindent{KEY WORDS:  Exact computation, Fast algorithm, Methylation levels imputation, Multiple functional data, Stochastic differential equations} 
\end{abstract}
\end{titlepage}

\section{Introduction}
\label{sec:intro_nonseparable}
 The increasing demands to analyze  high dimensional data with complex structures  have facilitated  the development of novel statistical models for functional data,  in which the outcomes can be interpreted as samples of random functions.  Time series, longitudinal, and spatial data are some typical examples of functional data. 
One common feature among functional data  is that, often, the linear regression does not appropriately explain the correlations between the outcomes that are close in the inputs of the function. The correlation is often expressed through a mapping from the functional inputs to the associated outcomes, usually modeled as a stochastic process, and the correlations between nearby inputs are captured through a covariance matrix. One natural choice of  such stochastic process is the Gaussian stochastic process (GaSP), which  has been widely  used in many applications \citep{sacks1989design,Bayarri09,gelfand2010handbook}.

 GaSP models have also been popular in analyzing functional data with multiple functional outcomes, in which independent GaSP models are  generally  built separately for each outcome for simplicity.   
A more sophisticated approach is to define a {separable} GaSP model, where  the correlations between functions and between inputs are modeled separately  
  using a  matrix normal distribution \citep{conti2010bayesian}.  Other approaches include the use of a linear model of coregionalization, where the factor processes modeled as independent GaSPs to explain the variation over the input space. This construction results in  \emph{nonseparable} covariance structures, meaning that the covariance matrix cannot be decomposed as a Kronecker product of two small covariance matrices.  

For  large-scale  data, a GaSP model is often computationally expensive:  the  evaluation of the likelihood requires $O(n^3)$ computational operations to compute the inverse of the covariance matrix, where $n$ is the number of observed data points.  {To ease the computation, } many approximation methods have been proposed, including low rank approximation \citep{banerjee2008gaussian}, covariance tapering  \citep{kaufman2008covariance}, use of Gaussian Markov random field representations \citep{lindgren2011explicit}, and likelihood approximation \citep{eidsvik2013estimation}. {Those approximation methods are sometimes preferred  for computationally intensive problems, however, the exact computation is more desired if we can overcome the $O(n^3)$ computational operations. }

 Our motivating study is to impute millions of DNA methylation levels at CpG sites across the human genome. DNA methylation is an epigenetic modification of DNA,  playing important roles in  DNA replication, gene transcription, aging, and cancer evolution \citep{das2004dna,scarano2005dna}.  Methylation levels are quantified 
at every genomic CpG site, a region of DNA where a cytosine (C) nucleotide is followed by a guanine (G) nucleotide in the linear sequence of bases along its 5' to 3' direction.  
Single-site DNA methylation level can be quantified by whole-genome bisulfite sequencing (WGBS), in which approximately 26 million 
CpG sites in the human genome are evaluated for whether they are methylated or not. 
However, WGBS is expensive and hard to examine in certain genomic regions. This motivates alternative methylation assay technologies,  such as  Illumina HumanMethylation450 BeadChip (henceforth, Methylation450K) that measures DNA methylation levels at approximately $482,000$ CpG sites (less than $2\%$ of the total number of CpG sites). The goal is to impute DNA methylation levels at the CpG sites that are observed in the WGBS samples but unobserved in the Methylation450K data by exploiting the correlations among the full set of CpG sites in the WGBS samples.

\begin{figure}[ht]
\centering

  \begin{tabular}{cc}
\raisebox{-3.1mm}[0pt][0pt]{
	\includegraphics[height=.36\textwidth,width=.5\textwidth]{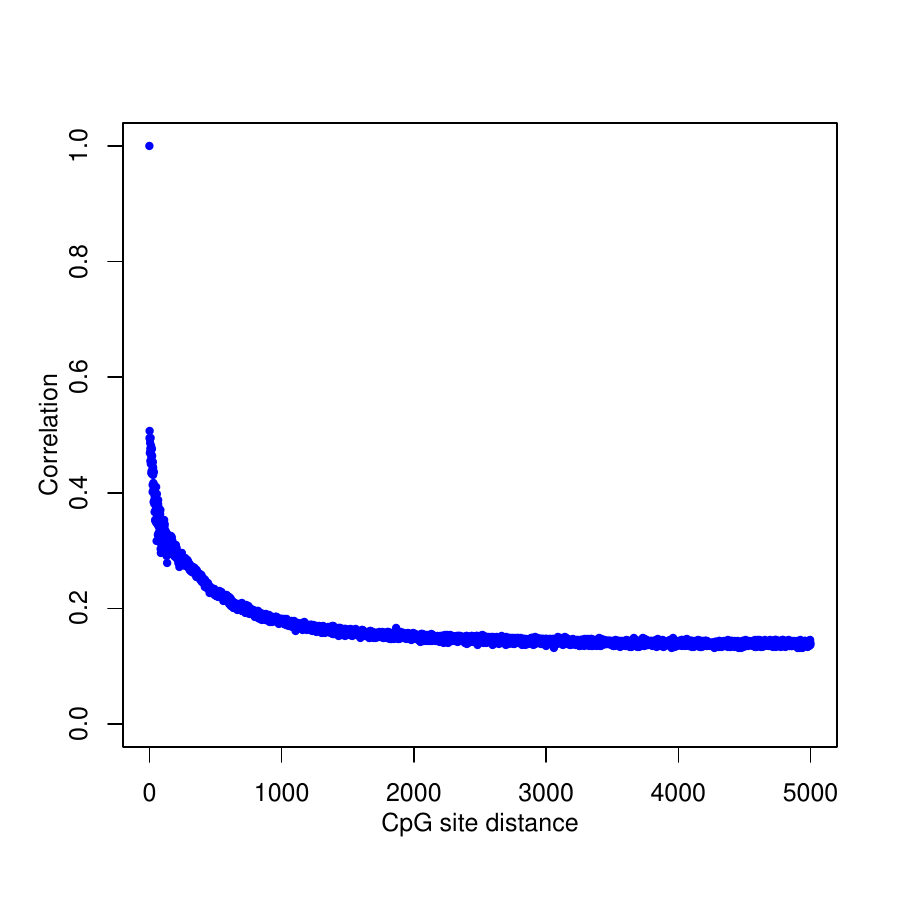}
	}
  \includegraphics[height=.3\textwidth,width=.5\textwidth]{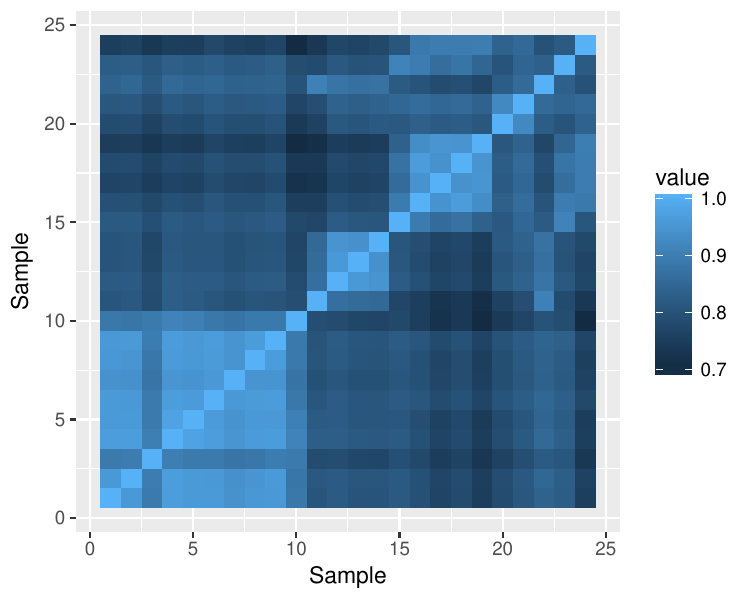} \\

  \end{tabular}

   \caption{Empirical correlation of methylation levels across sites (left panel) and across samples (right panel), based on 24 samples and one million methylation levels in chromosome 1 of each sample. }
\label{fig:empirical_corr}
\end{figure}

We first explore the empirical correlation of methylation levels  studied in the WGBS data \citep{ziller2013charting}. For each integer distance smaller than 5000 bases, we plot the empirical correlation of every possible pair with this distance in the left panel of Figure~\ref{fig:empirical_corr}. The blue dots are the average correlation of the methylation levels between two CpG sites at a given CpG distance smaller than 5000 bases. The methylation levels at nearby CpG sites are   correlated to each other on average and the correlation gradually decays as the distance between the two CpG sites increases. Such phenomenon is called co-methylation and has  been observed in previous studies \citep{zhang2015predicting}. 
 Furthermore, we plot the correlation of the methylation levels across samples in the right panel of Figure~\ref{fig:empirical_corr} \citep{ggplot2,reshape2}, which shows the methylation levels are highly correlated for biological reasons.  Since methylation plays important roles in suppressing gene expression levels, they are tightly regulated in cells and variability of such regulations is associated with disease risk \citep{das2004dna}. Understanding the correlation patterns in methylations levels is thus meaningful for reducing the risk of diseases. 
   These empirical findings motivate us to develop a statistical model to exploit the correlations of methylation levels across the unequally-spaced genome sites and across  different  samples for the goal of imputation.

  In this paper,  we develop a computationally efficient model to impute the methylation levels. Our contribution lies in both application and methodology. First, we propose to use a nonseparable GaSP model to integrate different correlation structures among the methylation levels across genome sites and across samples into a coherent model, while the previous regression method in \cite{zhang2015predicting} ignores the correlations across samples. To achieve this goal, we propose a computationally approach for the nonseparable model in \cite{higdon2008computer}. We first show that the marginal likelihood of this model can be written as a product of the multivariate normal distributions of the transformed output, and then we apply the forward filtering and backward smoothing algorithm for the GaSP with a Mat{\'e}rn kernel function \citep{hartikainen2010kalman}. In our algorithm, the computational operations and storage are linear to the number of observations without an approximation to the likelihood function. We  implement this algorithm in an R package on CRAN \citep{Gu2019fastgasp} based on the R software \citep{R}. Lastly, we show that the proposed nonseparable GaSP model is a general statistical framework that unifies the linear regression and separable GaSP models.

The rest of the paper is organized as follows. In Section~\ref{sec:models}, we study a class of nonseparable GaSP models. The closed form marginal likelihood and predictive distribution are derived for the imputation problem. In Section~\ref{sec:computation}, the computational strategy for this class of nonseparable GaSP models is introduced, for which the computation scales linearly in the number of inputs of the function without approximation. In Section~\ref{sec:unification}, we unify some other frequently used approaches, such as the linear regression and separable GaSP models, under the framework of the nonseparable GaSP models.  Numerical examples and comparisons  to alternative  methods are provided in Section~\ref{sec:numerical_nonseparable}. We conclude the paper with discussion and future extensions in Section~\ref{sec:conc}.

\section{Modeling multiple functional data}
\label{sec:models}

 Let $y_i(s_j)$ be the methylation level of the $i^{th}$ sample at the $j^{th}$ CpG site, recording the proportion of probes for a single CpG site that is methylated, for $i=1, \dots, K$, $j=1, \dots, N$. 
Define two groups of sites, $\mathbf s^{\mathscr D}=\{s^{\mathscr D}_1,...,s^{\mathscr D}_n\} $ and  $\mathbf s^{*}= \{s^{*}_1,...,s^{*}_{n^*}\}$, where the methylation levels of $\mathbf s^{\mathscr D}$ are observed for all $K$ samples and the methylation levels of $\mathbf s^{*}= \{s^{*}_1,...,s^{*}_{n^*}\}$ are only available for the first $k$ samples but not available for the last $k^*$ samples.  The total number of samples is $K=k+k^*$ and the total number of CpG sites is $N=n+n^*$.

For the first $k$ samples, methylation levels are measured at all CpG sites, meaning that we observe $\mathbf y(\mathbf s^{\mathscr D})_{[k\times n]}$ and $\mathbf y(\mathbf s^{*})_{[k\times n^*]}$. However, for the remaining $k^*$ samples, the methylation levels are only observed at a small subset of CpG sites, denoted as  $\mathbf y^*(\mathbf s^{\mathscr D})_{[k^*\times n]}$. The methylation levels at the remaining CpG sites ($\mathbf y^*(\mathbf s^{*})_{[k^*\times n^*]}$) of these samples are unknown. Our goal, then, is to interpolate the  unobserved  methylation levels of these $k^*$ samples using their observed methylation values at $n$ CpG sites  and the full methylation values from the other $k$ samples. In other words, we seek the predictive distribution of $\mathbf y^{*}({\mathbf s^*}) $ conditional on $ \mathbf y(\mathbf s^{\mathscr D})$, $\mathbf y(\mathbf s^{*})$ and   $\mathbf y^{*}({\mathbf s^{\mathscr D}})$.

The imputation of methylation levels across  the  whole genome is computationally challenging due to the large number of CpG sites.  In the full WGBS data set, there are about $2.8\times 10^{7}$ CpG sites; even in the smaller  Methylation450K data, there are roughly $4.5\times 10^5$ CpG sites,  creating computational challenges.  In contrast,  the number of samples we are working with is relatively small: $24$ samples in the WGBS data and $100$ samples in the Methylation450K data. The key advantage of our method is that the computation required for imputation scales linearly in terms of the number of CpG sites. 

Here we make several extensions of a class of GaSP models with the nonseparable structure, which has been used for modeling multivariate spatially correlated data and functional outputs  \citep{goulard1992linear,higdon2008computer}. First of all, we construct a flexible way to incorporate the correlations across samples and across sites for prediction, with closed form expression of the marginal likelihood and predictive distribution. These expressions enable us to establish the connection between this nonseparable model and other models, such as the linear regression and separable models, discussed in Section~\ref{sec:unification}. Furthermore, we introduce a computationally feasible approach to large-scale problems with inputs (CpG sites) up to a million without approximating  the likelihood function. This computational feasible approach depends on the explicit form of the likelihood in Lemma \ref{lemma:lik_nonseparable1}, which can be computed linearly to the number of samples. We further give an exact and fast computation in Section \ref{sec:computation} such that the overall computation of the likelihood is exact and at the order of $O(nK)$ operations, rather than $O((nK)^3)$ operations.  The proof of this section is given in Appendix. Without loss of generality, we assume the data are centered at zero. An extension to combine site-specified features is given  in the supplementary materials.

  \subsection{Nonseparable GaSP model}

 We start with the linear model of coregionalization (LMC) of the outputs $\mathbf Y(s)$, a $K\times 1$ vector for every site $s \in \mathscr{S}$ (\cite{goulard1992linear,gelfand2004nonstationary,higdon2008computer,chang2014fast}):
   \begin{align}
   \begin{split}
   \mathbf Y(s)& = \mathbf A \mathbf {\tilde v}(s)  +\bm \epsilon_0,\\
              \end{split}
   \label{equ:nonseparable1}
   \end{align}
where $\bm \epsilon_0 \sim \MN(\mathbf 0, \sigma^2_0\mathbf I_{K})$,  $\mathbf A=(\mathbf a_1;...; \mathbf a_K)$ is a $K \times K$ matrix  with  $\mathbf a_i$ being  the $i^{th}$ factor loading vector ($K \times 1$ vector) specified later and $\mathbf {\tilde v}(s)=(\tilde v_1(s),...,\tilde v_K(s))^T$ is a vector of random factors at site $s$. Note that one may define $\mathbf A$ to be a $K \times d$ matrix with $d\leq K$ for other applications, but since $K$ is small in our application, we do not pursue the dimension reduction toward this direction.

 As shown in Figure~\ref{fig:empirical_corr}, the correlation of the methylation levels at nearby CpG sites decays as the genomic distance increases.  This motivates us to model each random factor processes $\tilde v_i(\cdot)$ independently as a zero mean GaSP with noises for $s \in \mathscr{S}$:
  \begin{align}
     \begin{split}
  \tilde v_i(s)&=v_i(s)+ \epsilon_i, \\
  v_i(\cdot)&\sim \GaSP(0,  \tau^2_i c_i(\cdot, \cdot)), 
                \end{split}
  \label{equ:GaSP}
  \end{align}
  where $\epsilon_i \sim \N( 0, \sigma^2_i)$ is an independent noise and  $\tau^2_ic_i(s_a, s_b)$ is the covariance between any site $s_a, s_b \in \mathscr{S}$, with unknown  variances $\sigma^2_i$ and $\tau^2_i$, respectively. Denote the nugget variance ratio parameter $\eta_i=\sigma^2_i/\tau^2_i$. Form (\ref{equ:GaSP}) implies the marginal distribution of $\mathbf {\tilde v}_i(\mathbf s^{\mathscr D})=(\tilde v_i(s^{\mathscr D}_1),....,\tilde v_i(s^{\mathscr D}_n))^T$ follows a multivariate normal distribution below:
\[ \mathbf {\tilde v}_i(\mathbf s^{\mathscr D}) \mid  \tau^2_i, \eta_i  \sim  \MN(\mathbf 0, \tau^2_i(\mathbf R_i +\eta_i\mathbf I_n ) ), \]
  where the $(l, \, m)$ entry of the correlation matrix $\mathbf R_i$ is $ c_i(s^{\mathscr D}_l,  s^{\mathscr D}_m)$ for $i=1,...,K$, and $\mathbf I_n$ is an $n\times n$ identity matrix.  The Mat{\'e}rn kernel is often used for modeling the correlation:
 \begin{equation}
 c_i(d)=\frac{1}{2^{\nu_i-1}\Gamma(\nu_i)}\left(\frac{d}{\gamma_i} \right)^{\nu_i} \mathcal K_{\nu_i} \left(\frac{d}{\gamma_i} \right),
 \label{equ:matern}
 \end{equation}
 where  $\Gamma(\cdot)$ is the gamma function, $\mathcal{K}_{\nu_i}(\cdot)$ is the modified Bessel function of the second kind with a smoothness parameter $\nu_i$, $\gamma_i$ is a range parameter,  and $d=|s_a-s_b|$ for any $s_a, s_b \in \mathscr S$. {When $\nu_i= (2m_i+1)/2$ with $m_i \in \mathbb N$, the GaSP with a Mat{\'e}rn kernel is $m^{th}_i$ sample path differentiable and the Mat{\'e}rn kernel has a closed form expression in these scenarios. For instance, the  exponential kernel is equivalent to the Mat{\'e}rn kernel with $\nu_i=1/2$ and the Gaussian kernel is the Mat{\'e}rn kernel with $\nu_i \to +\infty$. The flexibility of the Mat{\'e}rn kernel makes it widely applicable for modeling spatially correlated data \citep{gelfand2010handbook}.  }  
 
 Denote ${ \mathbf Y(\mathbf s^{\mathscr D})=(\mathbf y(\mathbf s^{\mathscr D})^T;\mathbf y^*(\mathbf s^{\mathscr D})^T   )^T }$. Following \cite{higdon2008computer}, we apply the singular value decomposition (SVD) to $\mathbf Y(\mathbf s^{\mathscr D})=\mathbf U \mathbf D \mathbf V$ and estimate $\mathbf A$ as $\mathbf { A}=\mathbf U \mathbf D/\sqrt{n}$, for the following reasons. First of all, the computational order of estimating $\mathbf A$ is linear to $n$, which is essential when $n$ is at the size of $10^6$.  Secondly, we have $\mathbf a^T_i \mathbf a_j=0$ if $i\neq j$, and hence $\mathbf { A}^T \mathbf { A}$ is a diagonal matrix, which substantially simplifies the computation of the likelihood.     Moreover, $\mathbf { A} \mathbf { A}^T=\mathbf Y(\mathbf s^{\mathscr D}) \mathbf Y(\mathbf s^{\mathscr D})^T/n$, unifying the  linear regression model under the framework of the nonseparable model shown later in Remark \ref{remark:nonseparable_LM} of Section \ref{sec:unification}.  
 
Denote the $K\times n$ factor matrix $\mathbf {\tilde v}(\mathbf s^{\mathscr D})$ with the $(i,j)$ term being $\tilde v_i(s^{\mathscr D}_j)$ for $1\leq i\leq K$ and $1\leq j\leq n$. Note the $\mathbf{\tilde v}(\mathbf s^{\mathscr D})$ can be marginalized explicitly. We first vectorize the outputs $\mathbf  Y_{v}(\mathbf s^{\mathscr D}):=vec(\mathbf Y(\mathbf s^{\mathscr D}))$  and factor matrix  $\mathbf {\tilde v}_v(\mathbf s^{\mathscr D}):=vec(\mathbf{\tilde v}(\mathbf s^{\mathscr D})^T )$,  both of which are $Kn$-dimensional vectors.  Define a $Kn\times Kn$ matrix $\mathbf A_{v}:= [\mathbf I_{n}\otimes \mathbf a_1;...; \mathbf I_{n}\otimes \mathbf a_K]$. By simple algebra, model (\ref{equ:nonseparable1}) can be written as
 \begin{equation}
 \mathbf  Y_{v}(\mathbf s^{\mathscr D})= \mathbf A_{v} \mathbf{\tilde v}_v(\mathbf s^{\mathscr D})+\bm \epsilon_{0v}, 
 \end{equation}
   where $\bm \epsilon_{0v}\sim \MN(\mathbf 0, \sigma_0^2 \mathbf I_{nK})$ and $ (\mathbf {\tilde v}_v(\mathbf s^{\mathscr D})\mid   \tau^2_1,...,  \tau^2_K, \mathbf {\tilde R}_1,..., \mathbf{\tilde R}_K) \sim  {\MN}(\mathbf 0, \bm \Sigma_v)$, with the $(l,m)$ entry of the $\mathbf{\tilde R}_i$ being  $\tilde c_i(s^{\mathscr D}_l, s^{\mathscr D}_m)=c_i(s^{\mathscr D}_l, s^{\mathscr D}_m)+\eta_i 1_{l=m}$.  Here $\bm \Sigma_v =blkdiag(\tau_1^2 \mathbf {\tilde R}_1;...;\tau_K^2 \mathbf {\tilde R}_K) $, where $blkdiag(.)$ means the block diagonal matrix between sites.    As shown in \cite{higdon2008computer}, directly marginalizing out  $\tilde{\mathbf v}_v(\mathbf s^{\mathscr D})$ leads to:
    \begin{equation}
    \mathbf  Y_{v}(\mathbf s^{\mathscr D}) \mid  \sigma^2_0, \tau^2_1,..,  \tau^2_K, \mathbf {\tilde R}_1,..., \mathbf {\tilde R}_K \sim  {\MN}(\mathbf 0,  \mathbf A_{v} \bm \Sigma_v \mathbf A^T_{v} +\sigma^2_0 \mathbf I_{Kn} ).
    \label{equ:complicated_lik}
    \end{equation}
 The straightforward computation of the likelihood (\ref{equ:complicated_lik}), however, requires to invert a  $Kn\times Kn$   covariance matrix  $\mathbf A_{v} \bm \Sigma_v \mathbf A^T_{v} +\sigma^2_0 \mathbf I_{Kn} $, which takes $O( (Kn)^3 )$ operations in general.   The following lemma states the likelihood in (\ref{equ:complicated_lik}) can be computed  with $O(Kn^3)$ operations.
  \begin{lemma}
Assume $\mathbf { A}=\mathbf U \mathbf D/\sqrt{n}$, where $\mathbf U$ and $\mathbf D$ are  from the SVD of $\mathbf Y(\mathbf s^{\mathscr D})=\mathbf U \mathbf D \mathbf V$. After integrating out   $\mathbf v(\mathbf s^{\mathscr D})$, the marginal likelihood of $\mathbf  Y(\mathbf s^{\mathscr D})$ in model (\ref{equ:nonseparable1}) follows a product of $K$ independent multivariate normal distributions,
 \begin{equation*}
 \label{equ:marlik_nonseparable1}
 p(\mathbf  Y(\mathbf s^{\mathscr D} ) \mid \sigma^2_0, \tau^2_1,..,  \tau^2_K, \mathbf {\tilde R}_1,..., \mathbf {\tilde R}_K ) =|\mathbf A_v^T \mathbf A_v|^{-\frac{1}{2}}  \prod^{K}_{i=1} p_{ {MN}}(\hat {\mathbf  v}_i(\mathbf s^{\mathscr D}); \mathbf 0, \tau^2_i \mathbf{\tilde R}_i+ \sigma^2_0 (\mathbf a^T_i \mathbf a_i)^{-1} \mathbf I_{n}), 
 \end{equation*}
   where $p_{ {MN}}(.\,; {\bm \mu},{\bm \Sigma})$ denotes the multivariate normal density with mean $\bm \mu$ and covariance $\bm \Sigma$, $\hat {\mathbf v}_i(\mathbf s^{\mathscr D})$ is the transpose of the $i^{th}$ row of 
   \begin{equation}
   \hat {\mathbf v}(\mathbf s^{\mathscr D}) = (\mathbf A^T \mathbf A)^{-1}\mathbf A^T \mathbf Y(\mathbf s^{\mathscr D}). \vspace{-.4in} 
   \label{equ:v_hat_s_D}
   \end{equation} 
   \label{lemma:lik_nonseparable1}
  \end{lemma}
  Lemma~\ref{lemma:lik_nonseparable1} states that the marginal likelihood by model (\ref{equ:nonseparable1}) can be written as a product of $K$ multivariate normal densities, which simplifies the computation.   In particular, instead of computing the multivariate normal densities with a $Kn \times Kn$ covariance matrix, one can evaluate the densities by $K$ independent multivariate normal distributions, each of which has an $n\times n$ covariance matrix.   The direct computation of the inverse of an $n\times n$ covariance matrix, however, is still very hard in general, when $n$ is at the size of  $10^6$.  An efficient way that computes the exact likelihood will be  provided in Section~\ref{sec:computation}.    
  
Denote $\mathbf Y( s^*_j)=(\mathbf y( s^*_j)^T; \mathbf y^*( s^*_j)^T)^T$, where $\mathbf y( s^*_j)$ and $\mathbf y^*( s^*_j)$ are the $j^{th}$ column of $\mathbf y( \mathbf s^*)$ and  $\mathbf y^*( \mathbf s^*)$ respectively.  The goal of imputation  is to find the predictive distribution at an unexamined site $s^*_j$  conditioning on the available data, which is given in the lemma below. 

  \begin{lemma} 
  We assume the same conditions in Lemma~\ref{lemma:lik_nonseparable1}.
\begin{description}
\item[1.]  
For every $s^*_j$, $j=1,...,n^*$, denote $\mathbf  r_i(s^*_j)=\left(c_i(s^*_j, s^{\mathscr D}_1 ),...,c_i(s^*_j, s^{\mathscr D}_n )\right)^T$,  $\mathbf D^*(s^*_j)$. One has
 \[\mathbf Y(s^*_j) \mid \mathbf Y(\mathbf s^{\mathscr D}), \sigma^2_0, \bm \tau^2_{1:K}, \bm \gamma_{1:K}, \bm \eta_{1:K} \sim  \MN \left(\bm {\hat \mu}(s^*_j), \mathbf {\hat \Sigma}(s^*_j) \right).\]
 Here $\bm {\hat \mu}(s^*_j)=\mathbf A \mathbf {\hat v}^*(s^*_j)$,  where the $i$th entry of $\mathbf {\hat v}^*(s^*_j)$ is $\hat v^*_i(s^*_j)=\mathbf r_i^T(s^*_j) (\mathbf {\tilde R}_i+ \frac{\sigma^2_0 (\mathbf a^T_i \mathbf a_i)^{-1} }{ \tau^2_i} \mathbf I_n )^{-1} \hat {\mathbf v}_i(\mathbf s^{\mathscr D})$ with  $\hat {\mathbf v}_i(\mathbf s^{\mathscr D})$ being the transpose of the $i^{th}$ row of   $\hat {\mathbf v}(\mathbf s^{\mathscr D}) $ defined in Equation (\ref{equ:v_hat_s_D}), and $\mathbf {\hat \Sigma}(s^*_j)=\mathbf A \mathbf D^*(s^*_j) \mathbf A^T+\sigma^2_0 \mathbf I_{K}$, where   $\mathbf D^*(s^*_j)$ is a  diagonal matrix with the $i$th diagonal term being  $\tau^2_i( \tilde c_i(s^*_j, s^*_j)-\mathbf r_i^T(s^*_j) (\mathbf{\tilde R}_i+ \frac{\sigma^2_0 (\mathbf a^T_i \mathbf a_i)^{-1}}{ \tau^2_i}\mathbf I_n )^{-1} \mathbf r_i(s^*_j) )$, for $i=1,...,K$.
 
 \item[2.]  Denote the partition $\bm {\hat \mu}(s^*_j)=\left(\bm {\hat \mu}^T_0(s^*_j), \bm {\hat \mu}^T_*(s^*_j) \right)^T$ and $\bm {\hat \Sigma}(s^*_j)= \left( {\begin{array}{*{20}{c}}
\bm {\hat \Sigma}_{00} (s^*_j)& \bm {\hat \Sigma}_{0*}(s^*_j)\\
 \bm {\hat \Sigma}_{*0}(s^*_j)& \bm  {\hat \Sigma}_{**}(s^*_j)  \\
 \end{array} } \right).$
 \noindent For every $s^*_j$, the predictive distribution of the unobserved $\mathbf y^*(s^*_j)$ follows
\begin{equation*}
\mathbf y^*(s^*_j)\mid \mathbf y(\mathbf s^{\mathscr D}),  \mathbf y(\mathbf s^*), \mathbf y^*(\mathbf s^{\mathscr D}), \sigma^2_0, \bm \tau^2_{1:K}, \bm \gamma_{1:K}, \bm \eta_{1:K} \sim {\text{MN}} \left(\bm {\hat \mu}_{*|0}(s^*_j), \mathbf {\hat \Sigma}_{*|0}(s^*_j) \right),
\end{equation*}
 where $\bm {\hat \mu}_{*|0}(s^*_j)=\bm{\hat \mu}_{*}(s^*_j)+ \bm {\hat \Sigma}_{*0}(s^*_j)\bm {\hat \Sigma}^{-1}_{00}(s^*_j) \left( \mathbf y(s^*_j )-\bm {\hat \mu}_0 (s^*_j) \right)   $  and  $\bm {\hat \Sigma}_{*|0}(s^*_j)= \bm {\hat \Sigma}_{**}(s^*_j)  -\bm {\hat \Sigma}_{*0}(s^*_j)\bm {\hat \Sigma}^{-1}_{00}(s^*_j) \bm {\hat \Sigma}_{0*}(s^*_j)$.  
\end{description}
\label{lemma:pred_nonseparable1}
\end{lemma}

In the methylation levels imputation study, $\bm {\hat \mu}_{*|0}(s^*_j)$ can be used as predictions for  $\mathbf y^*(s^*_j)$ for any site $s^*_j$,  by properly conditional on all observations.  Note that the model and predictive distribution given in Lemma \ref{lemma:lik_nonseparable1} and Lemma \ref{lemma:pred_nonseparable1} are not identifiable between $\eta_1,...,\eta_K$, and $\sigma^2_0$.  In the following, we simply constrain $\sigma^2_0=0$ to avoid the potential identifiability issue.

\begin{figure}[t]
\centering

  \begin{tabular}{cc}
	\includegraphics[height=.3\textwidth,width=1 \textwidth]{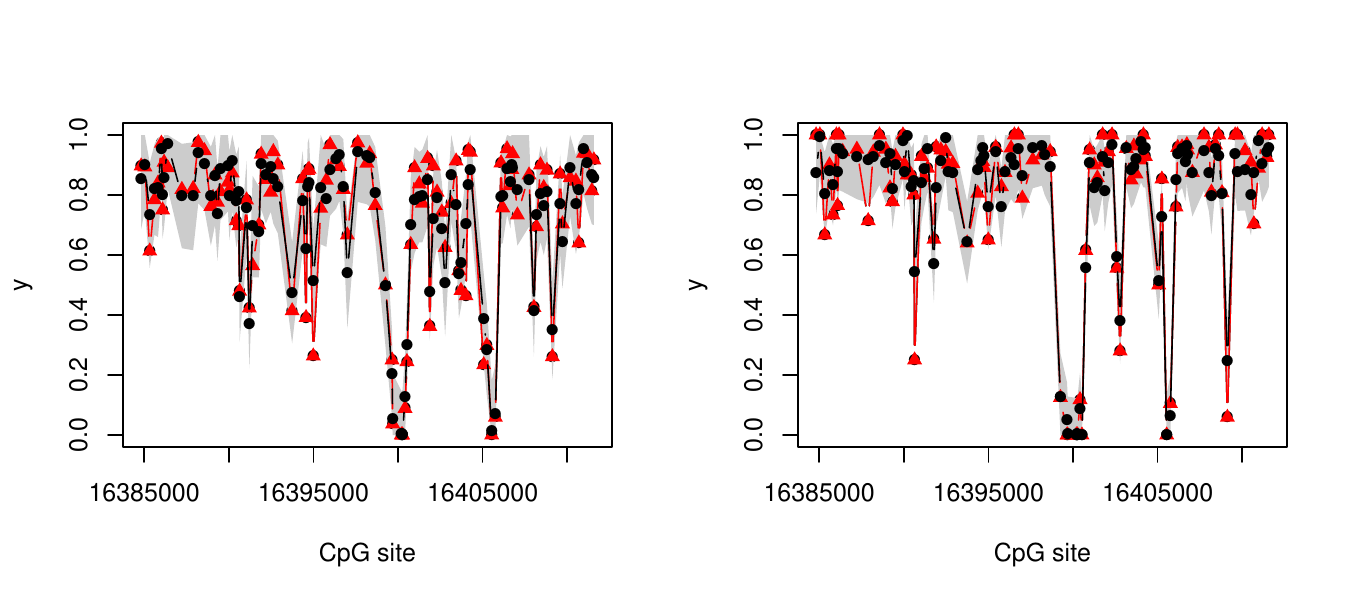} \vspace{-.5in}\\

    \includegraphics[height=.3\textwidth,width=1\textwidth]{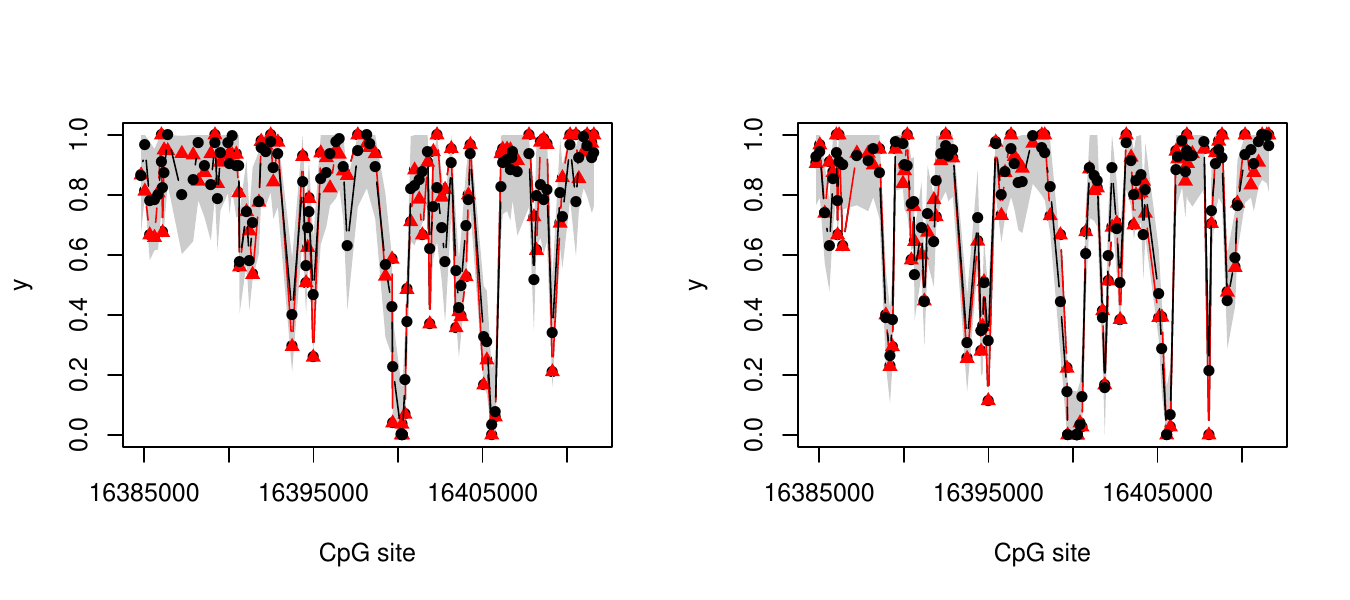}    
  \end{tabular}

   \caption{The held-out methylation levels (red triangles)  and  prediction of methylation levels  (black dots) by the nonseparable GaSP model for  randomly selected  4 samples at 100 CpG sites. The 95\% posterior predictive interval is  dashed as the shaded area. $25\%$ of the observations at the first million CpG sites in chromosone 1 in the WGBS data of these 4 samples are held out for testing. }
\label{fig:plot_test_ppl_4}
\end{figure}

Figure~\ref{fig:plot_test_ppl_4} plots the predicted methylation levels as the black dots  at 100 held-out CpG sites for 4 samples in the WGBS data,  and held out methylation levels are marked as the red triangles.     The prediction by the nonseparable GaSP model captures the pattern of the methylation levels reasonably well, with an adequate length of 95\% predictive interval, graphed as the shaded area.  { A more detailed comparison between the nonseparable GaSP model and other computational feasible alternatives is given in Section~\ref{sec:numerical_nonseparable}.}

\section{Computational strategy}
\label{sec:computation}
 The computations of the likelihood in model (\ref{equ:nonseparable1}) require to compute the inverse of $\mathbf R_i$, each with $O(n^3)$ operations, making the implementation impractical when $n$ is large.  We  introduce a computationally efficient algorithm, based on the connection between the Gaussian random field  and Gaussian Markov random field (GMRF) \citep{whittle1954stationary,whittle1963stochastic}.  The idea  was recently introduced in  \cite{hartikainen2010kalman}. Unlike many other methods, no approximation to the likelihood is needed in this approach. We briefly review this computational strategy and extend it to compute the likelihood of the nonseparable GaSP model.

Consider a continuous auto-regressive model with order $p$, defined by a stochastic differential equation (SDE),
\begin{equation}
c_p v^{(p)}(s)+c_{p-1}v^{(p-1)}(s)+...+c_0v(s)=b_0z(s), 
\label{equ:cont_state}
\end{equation}
where $v^{(l)}(s)$ is the $l^{th}$ derivative of $v(s)$ and $z(s)$ is the standard Gaussian white noise process defined on $s \in \mathbb R$.  Here we set $c_p=1$ to avoid the nonidentifiability issue.   The spectral density of equation (\ref{equ:cont_state}) is $S_{\mathbb R}(t)= \frac{b^2_0}{|C(2\pi \rm{i} t)|^2}$, where $\rm{i}$ is the imaginary number, and the operator $C(\cdot)$ is defined by $C(z)=\sum^p_{l=0}c_l z^l$. The form of the above spectral density is $
S_{\mathbb R}(t)=\frac{\mbox{constant}}{\mbox{polynomial  in} \,\, t^2}$, which is a rational functional form. It has been shown in \cite{whittle1954stationary,whittle1963stochastic} that the spectral density of GaSP with the Mat{\'e}rn covariance is 
\begin{equation}
S_{Mat}(t)\propto \frac{1}{(\lambda^2+t^2)^{(\nu+1/2)}},
\label{equ:matern_spectral}
\end{equation}
where $\lambda=\frac{\sqrt{2\nu}}{\gamma}$ with the range parameter $\gamma$ and the smoothness parameter  $\nu$.  The spectral density in (\ref{equ:matern_spectral}) follows a rational functional form, meaning that we can utilize the GMRF representation for computation, elaborated in the following subsection.
\vspace{-.2in}
\subsection{The computation by continuous time stochastic process}
\label{subsec:computation_SDE}

As shown in Lemma \ref{lemma:lik_nonseparable1}, the likelihood of $\mathbf Y(\mathbf s^{\mathscr D})$ in model (\ref{equ:nonseparable1})  is proportional to $K$ multivariate normal distribution at $\mathbf {\hat v}_i(\mathbf s^{\mathscr D})$, which is the transpose of the $i$th row of $ \mathbf {\hat v}(\mathbf s^{\mathscr D})=(\mathbf A^T \mathbf A)^{-1}\mathbf A^T \mathbf Y(\mathbf s^{\mathscr D})$. We thus focus on  $ \mathbf {\hat v}(\mathbf s^{\mathscr D})$ with $\sigma^2_0=0$ due to the identifiability reason. 

For demonstration purposes, we assume a Mat{\'e}rn kernel with $\nu=5/2$ for  the latent factor processes 
\begin{equation}
c(d)= \left(1+\frac{\sqrt{5}d}{\gamma}+\frac{5d^2}{3\gamma^2}\right)\exp\left(-\frac{\sqrt{5}d}{\gamma}\right), 
\label{equ:matern2_5}
\end{equation}
with $d=|s_a-s_b|$ for any $s_a, s_b \in \mathscr S$. The computational advantages introduced in this subsection hold for Mat{\'e}rn kernel with $\nu=(2m+1)/2$ for all $m\in \mathbb N$. One can also use the Mat{\'e}rn kernel with different smoothness parameter $\nu_i$ for $\mathbf {\hat v}_i(\mathbf s^{\mathscr D})$  to reflect the change of the smoothness in the trajectories of factor processes.

For any $s \in \mathscr S$, the nonseparable GaSP model in  (\ref{equ:nonseparable1})  can be represented as
\begin{align}
 \label{equ:GP2}
 \begin{split}
 \hat v_i(s) &= v_i(s) + \epsilon_i,  \\
 v_i(\cdot)&\sim \GaSP(0, \tau_i^2c_i(\cdot, \cdot) ), 
    \end{split}
 \end{align}
where  $\epsilon_i \sim  \N(0,  \sigma^2_i )$ being an independent noise.  Denote   ${ \bm \theta_i}(s):=( v_i(s),  v^{(1)} _i(s),  v^{(2)} _i(s))^T$, where $ v^{(l)}_i(s) $ is the $l^{th}$ derivative of $ v_i(s)$ with regard to  $s$, for $l=1,2$.   The GaSP model defined in (\ref{equ:GP2}) with the correlation  in (\ref{equ:matern2_5}) follows an SDE: $  \frac{d\bm {\theta}_i(s)}{ds}=\mathbf J_i\bm {\theta}_i (s)+\sqrt{q_i}\mathbf L z_i(s)  $, 
where $q_i=\frac{16}{3}\sigma_i^2\lambda_i^5$, $z_i(s)$ is a standard Gaussian white noise process and $\lambda_i=\sqrt{2\nu_i}/\gamma_i$, for $i=1,...,K$. The closed form expression of  $\mathbf J_i$ and $\mathbf L$ is given in the supplementary materials.   

Denote  $\mathbf F=(1,0,0)$.  The solution of the  SDE for model (\ref{equ:GP2}) can be represented explicitly as follows \citep{hartikainen2010kalman},
\begin{align}
\label{equ:ctdlm}
\begin{split}
\hat v_i(s_{j+1})&= \mathbf F\bm \theta_i(s_{j+1}) + \epsilon_i, \\
\bm \theta_i(s_{j+1})&=\mathbf G_i(s_j) \bm \theta_i(s_j) +\mathbf w_i(s_j),  
\end{split}
\end{align}
where $\mathbf w_i(s_j) \sim \MN(\mathbf 0, \mathbf W_i(s_j))$ with $\mathbf G_i(s_j)=e^{\mathbf J_i(s_{j+1}-s_j)}$ and $\mathbf W_i(s_j) =\int^{s_{j+1}-s_j}_0 e^{\mathbf J_i t} \mathbf L q_i \mathbf L^T e^{\mathbf J_i^T t} dt$ for $i=1,...,K$ and  $j=1,...,N-1$.  The stationary distribution at $\bm \theta_i$ is $\bm \theta_i(s_{1}) \sim  {\MN}(\mathbf 0, \mathbf W_i(s_{1}) )$,  with $\mathbf W_i(s_{1})= \int^{\infty}_{0} e^{\mathbf J_it}\mathbf Lq_i\mathbf L^Te^{\mathbf J^T_i t}dt.$  All $\mathbf G_i(s_j) $, $\mathbf W_i(s_j)$, $\mathbf  W_i(s_{1})$,  and the likelihood of $\bm \theta_i$ are given in the supplementary materials.

The following Lemma \ref{lemma:KF_continuous_state_space} is the Kalman filter for the continuous state space model in (\ref{equ:ctdlm}) (see e.g. Chapter 4 in \cite{West1997} and Chapter 2 in \cite{petris2009dynamic}). 

\begin{lemma}[Kalman filter] 
\label{lemma:KF_continuous_state_space}
Consider  the continuous time state space model specified by Equation (\ref{equ:ctdlm}). To simplify the denotation, we assume all the distributions in this lemma are conditional on the parameters $(\sigma^2_i,\tau^2_i, \gamma_i)$. For each $i=1,...,K$, denote 
\[\bm \theta_i(s^{\mathscr D}_{j-1}) \mid \hat{\mathbf  v}_i(\mathbf s^{\mathscr D}_{1:j-1}) \sim  \MN( \mathbf m_i (s^{\mathscr D}_{j-1}), \mathbf C_i(s^{\mathscr D}_{j-1}) ), \]
for any $j\geq 2$.   One has the following statements for each $i=1,...,K$.
\begin{itemize}
\item[(i)] The one-step-ahead predictive distribution of $\bm \theta_i(s^{\mathscr D}_j)$ given  $\mathbf{\hat v}_i(s^{\mathscr D}_{1:j-1})$ is 
 \begin{equation}
 \bm \theta_i(s^{\mathscr D}_j)\mid \hat{\mathbf v}_i(s^{\mathscr D}_{1:j-1}) \sim  \MN(\bm \mu_{\bm \theta_i}(s^{\mathscr D}_j), \bm \Sigma_{\bm \theta_i}(s^{\mathscr D}_j)), 
 \label{equ:KF1}
 \end{equation}
where $\mu_{\bm \theta_i}(s^{\mathscr D}_j)=\E[\bm \theta_i(s^{\mathscr D}_j)\mid \hat{\mathbf  v}_i(\mathbf s^{\mathscr D}_{1:j-1}) ]= \mathbf G_i(s^{\mathscr D}_j) \mathbf m_i (s^{\mathscr D}_{j-1}) $ and 
 $   \bm \Sigma_{\bm \theta_i}(s^{\mathscr D}_j)=\Var[\bm \theta_i(s^{\mathscr D}_j)\mid  \hat{\mathbf  v}_i(\mathbf s^{\mathscr D}_{1:j-1}) ]= \mathbf G_i(s^{\mathscr D}_j) \mathbf C_i(s^{\mathscr D}_{j-1})  \mathbf G^T_i(s^{\mathscr D}_j)+\mathbf W_i(s^{\mathscr D}_j). $
 \item[(ii)] The one-step-ahead predictive distribution of $\hat{v}_i(s^{\mathscr D}_j)$ given $\hat{\mathbf  v}_i(\mathbf s^{\mathscr D}_{1:j-1})$ is  
 \begin{equation}
\hat{v}_i(s^{\mathscr D}_j) \mid \hat{\mathbf  v}_i(\mathbf s^{\mathscr D}_{1:j-1}) \sim  \N(f_i(s^{\mathscr D}_j) , Q_i(s^{\mathscr D}_j)  ),
 \label{equ:KF2}
\end{equation}
where     $f_i(s^{\mathscr D}_j)=\E[  \hat{ v}_i(\mathbf s^{\mathscr D}_j)   \mid \hat{\mathbf  v}_i(\mathbf s^{\mathscr D}_{1:j-1})]= \mathbf F \bm \mu_{\bm \theta_i}(s^{\mathscr D}_j)$ and     $Q_i(s^{\mathscr D}_j)=\Var[ \hat{ v}_i(\mathbf s^{\mathscr D}_j) \mid \hat{\mathbf  v}_i(\mathbf s^{\mathscr D}_{1:j-1}) ]= \mathbf F \bm \Sigma_{\bm \theta_i}(s^{\mathscr D}_j)  \mathbf F^T+ \sigma^2_i$. 
\item[(iii)] The filtering distribution of $\bm \theta_i(s^{\mathscr D}_j)$ given $\mathbf{\hat v}_i(s^{\mathscr D}_{1:j})$ is 
 \begin{equation}
\bm \theta_i(s^{\mathscr D}_{j})\mid \mathbf{\hat v}_i(s^{\mathscr D}_{1:j}) \sim \MN( \mathbf m_i (s^{\mathscr D}_j),  \mathbf C_i(s^{\mathscr D}_j)), 
 \label{equ:KF3}
\end{equation}
where  $  \mathbf m_i (s^{\mathscr D}_j)=\E[\bm \theta_i(s^{\mathscr D}_j)\mid \mathbf{\hat v}_i(s^{\mathscr D}_{1:j}) ]= \bm \mu_{\bm \theta_i}(s^{\mathscr D}_j) + \bm \Sigma_{\bm \theta_i}(s^{\mathscr D}_j)  \mathbf F^T  Q^{-1}_i(s^{\mathscr D}_j) e_i(s^{\mathscr D}_j)$  and  $\mathbf C_i(s^{\mathscr D}_j)=\Var[\bm \theta_i(s^{\mathscr D}_j)\mid \mathbf{\hat v}_i(s^{\mathscr D}_{1:j}) ]= \bm \Sigma_{\bm \theta_i}(s^{\mathscr D}_j) -  \bm \Sigma_{\bm \theta_i}(s^{\mathscr D}_j) \mathbf F^T   Q^{-1}_i(s^{\mathscr D}_j) \mathbf F \bm \Sigma_{\bm \theta_i}(s^{\mathscr D}_j)$, with $e_i(s^{\mathscr D}_j)=\hat{v}_i( s^{\mathscr D}_{j})  -f_i(s^{\mathscr D}_j)$ being the forecast error. 
\end{itemize}

\end{lemma}
  
  For the purpose of computing the likelihood discussed in the next subsection, we define the Kalman filter on $\mathbf s^{\mathscr D}$ in Lemma \ref{lemma:KF_continuous_state_space}.  Given an estimate of $(\sigma^2_i,\tau^2_i, \gamma_i)$ for $i=1,...,K$, we implement the Kalman filter for all sites $\mathbf s_{1:N}$,  which includes $\mathbf s^*$.  Filtering sites without an observation can be done in Kalman filter in Lemma \ref{lemma:KF_continuous_state_space} and we refer the reader to Section 2.7.3. in \cite{petris2009dynamic} for the details. The following Kalman smoother provides the predictive distribution for all sites $\mathbf s_{1:N}$  (see  \cite{West1997} and \cite{petris2009dynamic}).  
 
\begin{lemma}[Kalman smoother] 
Consider  the continuous time state space model specified by Equation (\ref{equ:ctdlm}).  We assume all the distributions in this lemma are conditional on  $( \sigma^2_i,  \tau^2_i,  \gamma_i)$. For $i=1,...,K$ and $j=1,...,N-1$, let $\bm \theta_i(s_{j+1} ) \mid \hat{\mathbf  v}_i(\mathbf s^{\mathscr D} ) \sim \mathcal \N( \bm \mu^*_{\bm \theta_i}(s_{j+1}) , \bm \Sigma^*_{\bm \theta_i}(s_{j+1}))$, then 
 \[\bm \theta_i(s_{j} )\mid \hat{\mathbf  v}_i(\mathbf s^{\mathscr D}) \sim \mathcal \MN( \bm \mu^*_{\bm \theta_i}(s_{j}), \bm \Sigma^*_{\bm \theta_i}(s_{j}) ), \]
 where $ \mu^*_{\bm \theta_i}(s_{j})= \mathbf m_i (s_j) +  \mathbf C_i(s_j) \mathbf G_i^T(s_{j+1}) \bm \Sigma^{-1}_{\bm \theta_i}(s_{j+1}) (\bm \mu^*_{\bm \theta_i}(s_{j+1})-\bm \mu_{\bm \theta_i}(s_{j+1}) )$ and
$ \bm \Sigma^*_{\bm \theta_i}(s_{j})  = \mathbf C_i(s_j)- \mathbf C_i(s_j) \mathbf G_i^T(s_{j+1}) \bm \Sigma^{-1}_{\bm \theta_i}(s_{j+1})  ( \bm \Sigma_{\bm \theta_i}(s_{j+1}) -\mathbf \bm \Sigma^{*}_{\bm \theta_i}(s_{j+1}) )\bm \Sigma^{-1}_{\bm \theta_i}(s_{j+1}) \mathbf G_i(s_{j+1}) \mathbf C_i(s_j)$.
\label{lemma:smoother}
\end{lemma}

\begin{figure}[t]
\centering
  \begin{tabular}{c}
  	\includegraphics[height=.35\textwidth,width=1\textwidth]{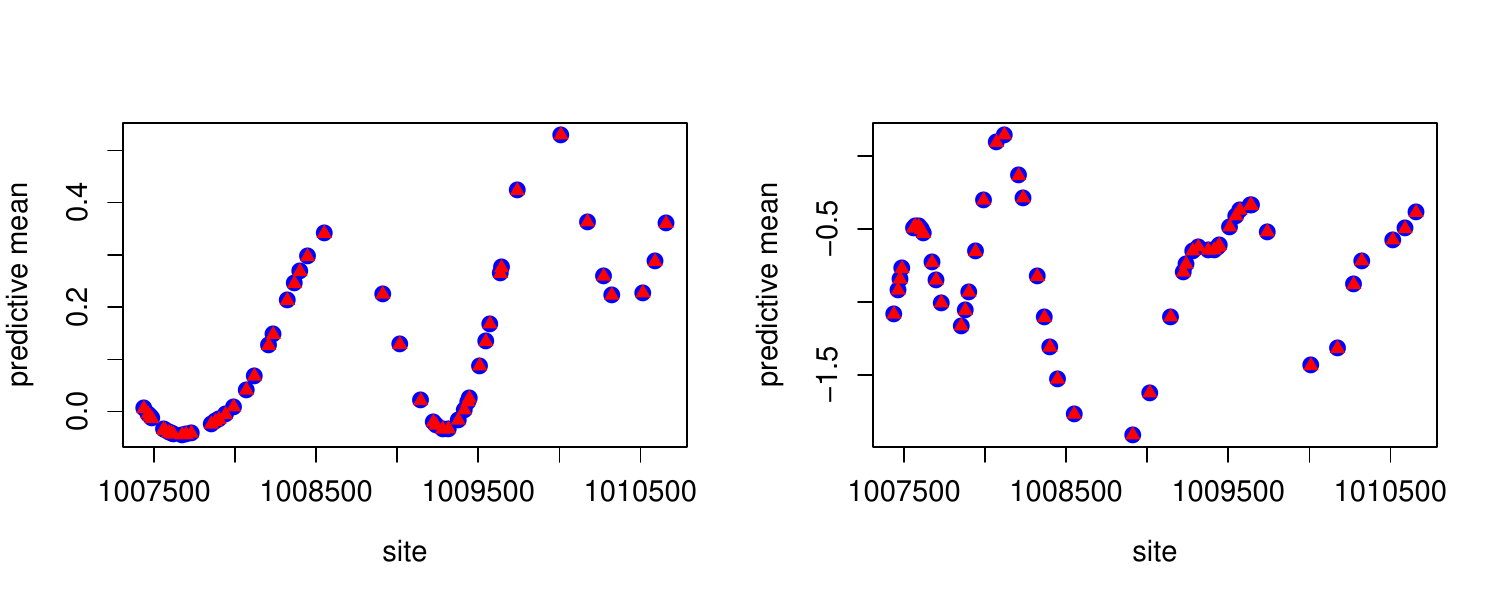} \vspace{-.15in}
  \end{tabular}
   \caption{Predictive mean from the direct inversion of the covariance and from the FFBS algorithm.   The blue dots are  $\mathbf {\hat v}_i(\mathbf s^*_{1:50}) |\mathbf {\hat v}_i(\mathbf s^{\mathscr D}_{1:400} )$ computed by equation (\ref{equ:ctdlm}) for $i=1$ (left panel) and $i=2$ (right panel), for a given set of parameters $( \sigma^2_i,   \gamma_i,\tau_i)$. The red triangles (overlapping the blue dots) are the same quantities by the direct computation for the GaSP model.  The root of mean square errors (RMSE) between the blue dots and red triangles at these 50 sites are $8.86\times 10^{-30}$ and $2.05 \times 10^{-30}$ for the left panel and right panel, respectively. }
\label{fig:prec}
\end{figure}

With the above setup,  the posterior  for  $\bm \theta_i(s)$, $i=1,..,K$, can be computed by a forward filtering and backward smoothing (FFBS) algorithm \citep{West1997}, which  only requires $O(n)$ computational operations, a lot smaller than $O(n^3)$ operations in the direct computation of the GaSP model. The prediction at $\mathbf s^*$ also only requires linear computational operations to the number of sites, so the total computational operations are only $O(N)$ altogether. The fast algorithm of evaluating the likelihood and making prediction is implemented in an R package on CRAN \citep{Gu2019fastgasp}.

  \begin{figure}[t]
\centering

  \begin{tabular}{c}
  	\includegraphics[height=.35\textwidth,width=1\textwidth]{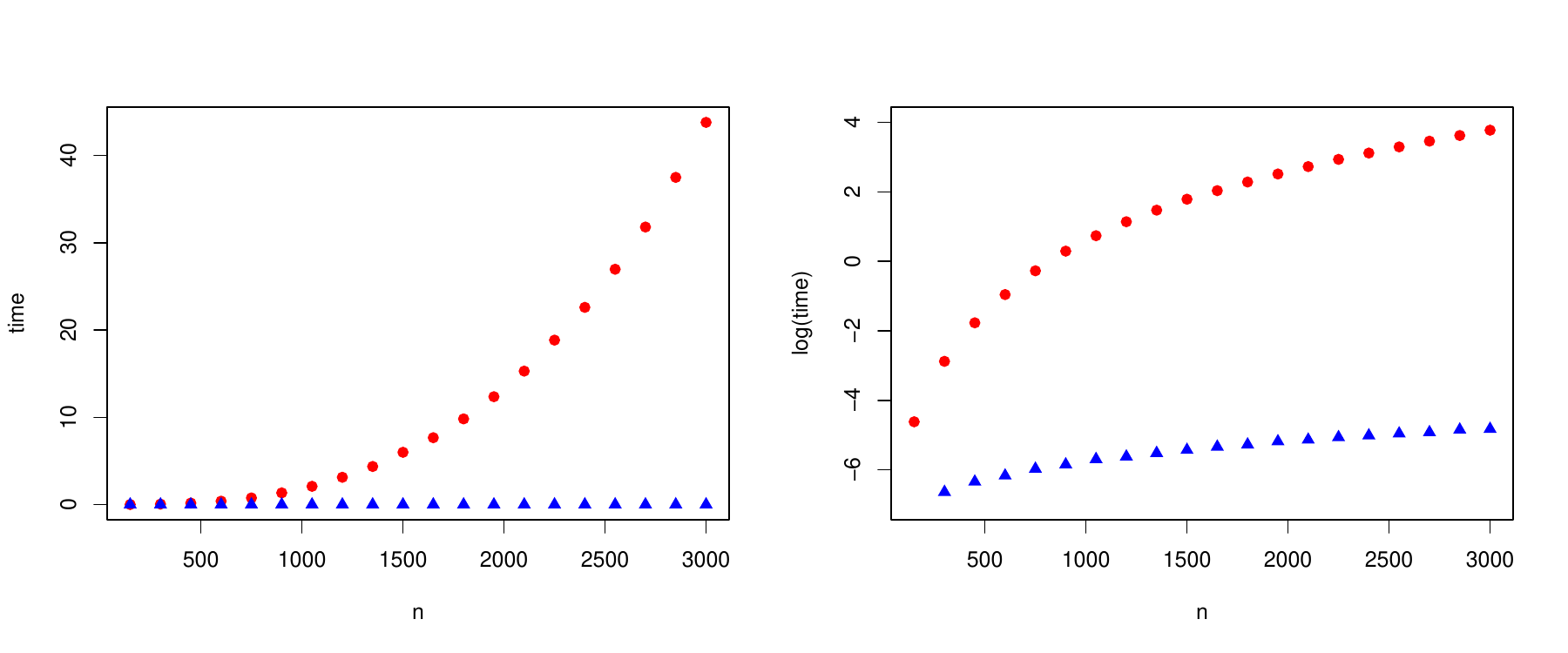}\vspace{-.15in}
  \end{tabular}
   \caption{Computational time in seconds  for one evaluation of the likelihood in seconds (left panel) and in natural logarithm of seconds (right panel).  The red dot represents the computational time by the direct computation and blue solid triangle is by the FFBS algorithm. }
\label{fig:time}
\end{figure}

  Figure~\ref{fig:prec} compares the posterior mean at some site $s^*_j$  by the FFBS algorithm and direct computation of GaSP  
 for a given set of parameters. Since they are  the same quantities computed in two different ways, the difference is extremely small. The main advantage of  our method  is that all summary statistics of interest, such as the posterior predictive mean and variance, as well as the marginal likelihood can be computed exactly. 

  The computational time between the algorithms, however,  differs significantly.   As shown in  Figure~\ref{fig:time}, the computation by the FFBS algorithm is a lot more efficient than the direct  evaluation of the likelihood, which requires $O(n^3)$ for matrix inversion. For instance, when $n=3,000$, evaluating the likelihood by  the FFBS algorithm takes less than 0.01 second, while the direct computation takes more than 40 seconds in R.   
  
   Note  when $\nu_i=5/2$, the Mat{\'e}rn covariance matrix in (\ref{equ:matern2_5}) and its inversion are both dense $n\times n$ matrices   with rank $n$.   However, the covariance matrix of the latent states, ${ \bm \theta_i}(s)=( v_i(s),  v^{(1)} _i(s),  v^{(2)} _i(s))^T$ is sparse, as  shown in supplementary materials. The result holds  for all Mat{\'e}rn classes when $\nu_i=(2m+1)/2$, with $m\in \mathbb N$. 

\subsection{Parameter estimation} 
\label{subsec:prior}

The most computationally intensive part of the FFBS algorithm discussed in Section \ref{subsec:computation_SDE}  is to sample the latent states.  Fortunately, one can marginalize out the latent states explicitly, 
\begin{align*}
p(\mathbf{ \hat {v}}( \mathbf s^{\mathscr D}) | \bm \sigma^2_{1:K}, \bm \tau^2_{1:K}, \bm \gamma_{1:K} ) =\prod^{K}_{i=1} \left\{p(\hat {v}_i(s^{\mathscr D}_1)|  \sigma^2_i,  \tau^2_i,  \gamma_i )\prod^{n}_{j=2} p(\hat {v}_i(s^{\mathscr D}_j)|\hat {v}_i(\mathbf s^{\mathscr D}_{1:j-1}), \sigma^2_i, \tau^2_i, \gamma_i )\right\},
\end{align*}
 each term of which follows a normal distribution given in the one-step look ahead prediction in the FFBS algorithm \citep{West1997}.  The  Lemma \ref{lemma:KF_lik} below show that the marginal likelihood can be computed with $O(n)$ computational operations.

  \begin{lemma}[Likelihood by the Kalman filter]
\label{lemma:KF_lik}
For each $i=1,...,K$, one has 
\begin{equation}
p(\hat{\mathbf v}_i(\mathbf s^{\mathscr D}_{1:j} ) \mid \sigma^2_i, \tau^2_i, \gamma_i)=(2\pi)^{-n/2}\prod^n_{j=1}Q^{-1/2}_i(s^{\mathscr D}_j) \exp\left\{-\sum^{n}_{j=1}\frac{ (\hat{v}_i( s^{\mathscr D}_{j})  -f_i(s^{\mathscr D}_j))^2 }{Q_i(s^{\mathscr D}_j)} \right\},
\label{equ:KF_lik}
\end{equation}
where $f_i(s^{\mathscr D}_j)$ and $  Q_i(s^{\mathscr D}_j)$ are given in Lemma \ref{lemma:KF_continuous_state_space}. 
\end{lemma}

Lemma \ref{lemma:KF_lik} is a direction consequence of the step (ii) in Lemma \ref{lemma:KF_continuous_state_space} so the proof is omitted here. To  complete  the model, we assume the prior as follows,
\begin{equation}
\pi(\bm \sigma^2_{1:K}, \bm \tau^2_{1:K}, \bm \gamma_{1:K})\propto \prod^{K}_{i=1} \frac{\pi( \tau^2_i,  \gamma_i ) }{\sigma^2_i}.
\label{equ:prior} 
\end{equation}
Denote $\zeta_i:=1/\gamma_i$ and $\eta_i:=\tau_i/\sigma^2_i$.   We assume the jointly robust prior \citep{gu2018jointly} for the transformed parameters $\pi(\zeta_i, \eta_i )\propto (C_i \zeta_i+\eta_i)^{a_i} exp(-b_i (C_i \zeta_i+\eta_i) )$,  for $i=1,...,K$, where $C_i$, $a_i$ and $b_i$ are prior parameters.  This prior approximates the reference prior in tail rates and is robust for posterior mode estimation \citep{Gu2018robustness}. { For the prior parameter, we use the default choice $a=1/2$, $b=1$, and $C_i=|\mathscr S|/n$, where $|\mathscr S|$ is the length of $\mathscr S$.}

As $n$ is  large,  the Markov chain Monte Carlo (MCMC) algorithm is still very slow for sampling the posterior distribution. We estimate the parameters by the posterior mode,
\begin{equation}
 (\hat \zeta_i, \hat \eta_i) =\mathop{argmax}_{\zeta_i, \eta_i} \left\{ p(\mathbf{ \hat {v}}( \mathbf s^{\mathscr D}) | \zeta_i,\eta_i ) \pi(\zeta_i,\eta_i)\right\}. 
 \label{equ:max_marginal_post}
 \end{equation} 

We summarize the estimation and prediction in  the nonseparable GaSP model in  Algorithm \ref{algorithm:0}. We assume the mean of each row of the observations were subtracted before the estimation and added back for predictions.

  \begin{algorithm}[t]
\caption{Estimation and prediction by the nonseparable GaSP model}
\raggedright
(1) Calculate $\mathbf A=\mathbf U \mathbf D/\sqrt{n}$, where $\mathbf U$ and $\mathbf D$ are from the SVD of $\mathbf Y(\mathbf s^{\mathscr D})$.

(2) Compute $ \mathbf {\hat v}(\mathbf s^{\mathscr D})=(\mathbf A^T \mathbf A)^{-1}\mathbf A^T \mathbf Y(\mathbf s^{\mathscr D})$. 

(3) Estimate the parameters by maximizing the posterior distribution in Equation (\ref{equ:max_marginal_post}), where the likelihood is calculated  by Equation (\ref{equ:KF_lik}).

(4) Plugging the estimates $(\hat \sigma^2_i, \hat \tau^2_i, \hat \gamma_i)$ for $i=1,...,K$ and implement the Kalman filter algorithm on all sites $\mathbf s_{1:N}$ using Lemma \ref{lemma:KF_lik}. The output is the predictive mean and predictive variance of $\hat{\mathbf v}_i(\mathbf x^*_j) \mid  \hat{\mathbf v}_i(\mathbf x^{\mathscr D}) $ for $j=1,...,n^*$.

(5) Produce the predictive distribution using Lemma \ref{lemma:pred_nonseparable1}. 

\label{algorithm:0}
\end{algorithm}

\section{Model unification}
\label{sec:unification}
In this section, we show the linear regression that was used for methylation level imputation \citep{zhang2015predicting}, and the separable GaSP model used in computer model emulation (\cite{conti2010bayesian,Gu2016PPGaSP}) are both special cases of the nonseparable GaSP model introduced in Section~\ref{sec:models}.
 
\subsection{Linear regression}
\label{subsec:two_LM}
 Assuming that  the samples are independent  to each other, a simple model is  to apply the linear regression separately for each CpG site $s_j$, $j=1,...,N$,  as follows,
 \begin{equation}
 y_i(s_{j})= \mathbf H_i(s_j) \bm \beta_j + \epsilon_{ij}, \, i=1,...,K,
\label{equ:regress_site}
 \end{equation}
 where $\mathbf H_i(s_j)$ are  the covariates for the $j^{th}$ CpG site of the $i^{th}$ sample and   $\epsilon_{ij} \sim N(0,\sigma^2_{0j})$ is an independent mean-zero Gaussian noise.  In \cite{zhang2015predicting}, some site-specific features, such as  methylation levels at nearby CpG sites, are used as covariates for imputation.   This approach assumes that methylation levels are independent across samples at every CpG site.  However,  methylation levels of different samples at a CpG site are generally correlated, as shown in Figure~\ref{fig:empirical_corr}. Numerical results will be shown that exploiting the correlation between samples leads to drastic improvement in imputation in Section~\ref{sec:numerical_nonseparable}.  
 
Alternatively, a regression model that exploits the correlations across samples follows
  \begin{equation}
y^*_i(s_j)=\mathbf H_i(s_j)\bm \beta_i +\epsilon_{ij},  \, j=1,...,N,  
  \label{equ:regress_ppl}
  \end{equation}
  with $\epsilon_{ij} \sim N(0,\sigma^2_{i0})$.  Here each site is treated independently. 
  Assume the methylation levels were first centered to have a zero mean and let $\mathbf H_i(s_j)= (y_1(s_j),...,y_k(s_j))$, meaning only the methylation levels of the $k$ samples with full observations are used as covariates, the model for $\mathbf s^{\mathscr D}$ can be expressed $\mathbf y^*_i(\mathbf s^{\mathscr D})^T= \mathbf y(\mathbf s^{\mathscr D})^T\bm \beta_i+\bm \epsilon_i, $  where $\mathbf y^*_i(\mathbf s^{\mathscr D})$ is the $i^{th}$ row of $\mathbf y^*(\mathbf s^{\mathscr D})$ and $\bm \epsilon_i \sim \MN(\mathbf 0, \sigma^2_{i0} \mathbf I_n)$.  The least squares (LS) estimator of $\bm \beta_i$ is $\hat {\bm \beta_i}= \left\{\mathbf y(\mathbf s^{\mathscr D}) \mathbf y(\mathbf s^{\mathscr D})^T \right\}^{-1}\mathbf y(\mathbf s^{\mathscr D}) \mathbf y_i^*(\mathbf s^{\mathscr D})^T$, where  $\mathbf y_i^*(\mathbf s^{\mathscr D})=\left(y_i^*(\mathbf s^{\mathscr D}_1),...,y_i^*(\mathbf s^{\mathscr D}_n)\right)$ being the $i^{th}$ row of $\mathbf y^*(\mathbf s^{\mathscr D})$.  The predictive mean of the methylation level of the $i^{th}$ sample at the $j^{th}$ unexamined site, $ \mathbf H_i(s^*_j) \hat {\bm \beta_i}$, follows
 \begin{align}
\E[\mathbf y^*_i(s^*_j )\mid \mathbf y(\mathbf s^{\mathscr D}),\mathbf y^*(\mathbf s^{\mathscr D}), \mathbf y(\mathbf s^{*}), \hat {\bm \beta_i} ] = \mathbf y( s^*_j)^T \left\{\mathbf y(\mathbf s^{\mathscr D}) \mathbf y(\mathbf s^{\mathscr D})^T \right\}^{-1}  \mathbf y(\mathbf s^{\mathscr D})  \mathbf y^*_i(\mathbf s^{\mathscr D})^T,
 \label{equ:pred_regress_ppl}
 \end{align}
 with  $\mathbf y( s^*_j)=(y_1(s^*_j),...,y_k(s^*_j))^T$ being the $j^{th}$ column of $\mathbf y( \mathbf s^*)$, for $i=1,...,k^*$, $j=1,...,n^*$. 

The following remark establish the connection between the linear regression and nonseparable GaSP model.

\begin{remark}
The predictive mean  under the linear regression model in (\ref{equ:pred_regress_ppl}) is identical to the predictive mean $\bm {\hat \mu}_{*|0}(s^*_j)$ of the nonseparable model with $\sigma^2_0=0$ in Lemma~\ref{lemma:pred_nonseparable1}  if in model (\ref{equ:nonseparable1}),
\item[(i.)] $\mathbf A=\mathbf U \mathbf D/\sqrt{n}$, where $\mathbf U $ and $\mathbf D$ are defined through the SVD of $\mathbf Y(\mathbf s^{\mathscr D})$.
\item[(ii.)] $\tilde v_i(\cdot)$ is the realization of independent mean zero Gaussian noise with the same variance.
 \label{remark:nonseparable_LM}
\end{remark}

\vspace{-.5in}
 \subsection{Separable model}
 \label{subsec:separableGaSP}
Denote $\mathbf Y$ as a  $K\times N$ matrix where each row is a sample of methylation levels at $N$ CpG sites.
 One may model the data through a joint model below 
   \begin{equation}
 \mathbf Y=\mathbf Z+\bm \epsilon_0,
 \label{equ:separable_model}
 \end{equation}
 where  $\bm \epsilon_0 $ is a zero-mean independent noise and $\mathbf Z$ is a  $K\times N$ random matrix modeled from a matrix-variate normal distribution, $\mathbf Z \sim {N}_{K,N}  (\bm \mu, \bm \Sigma, \mathbf R)$, with  a $K\times N$ mean matrix $\bm \mu$, a $K\times K$  row covariance matrix $\bm \Sigma$, and an $N\times N$ column correlation matrix $\mathbf R$.

 We call model (\ref{equ:separable_model}) \emph{the separable model}, as the correlations across samples and across sites are expressed separately by $\bm \Sigma$ and $\mathbf R$ respectively. Assuming that $\Lambda$ is modeled by a correlation function, where the $(i, \, j)$ entry $\mathbf R_{i,j}=c(d)$, with $c(\cdot)$ being the correlation function.
 The following remark states the separable  model defined in (\ref{equ:separable_model}) is a special case of the nonseparable model in (\ref{equ:nonseparable1}).

\begin{remark}
Assuming  $\bm \mu=\bm 0$, the separable model defined in  (\ref{equ:separable_model}) is equivalent to the nonseparable GaSP model in (\ref{equ:nonseparable1}) if 
 \item[(i.)] $\mathbf A$ is chosen such that $\bm \Sigma=\mathbf A \mathbf A^T$.
 \item [(ii.)] The covariance in (\ref{equ:GaSP}) has a unit variance, $c_i(d)=c(d)$, $\eta_i=0$, for $i=1,...,K$.
 \label{remark:nonseparable_separable}
\end{remark}

\section{Numerical  comparison}
\label{sec:numerical_nonseparable}
  We evaluate  the nonseparable GaSP model in (\ref{equ:nonseparable1}) and compare to  several alternative methods: two linear regression strategies (by site in (\ref{equ:regress_site}) and by sample in (\ref{equ:regress_ppl})), nearest neighbor method (using only the observed methylation level  closest to the unobserved site for prediction) and two  regression strategies by the random forest algorithm \citep{breiman2001random,randomforestR}.  We also introduce a localized Kriging method by partitioning the data into small blocks, and compare it with the nonseparable GaSP model in the supplementary materials.
 
The performance of these methods is evaluated on the out of sample prediction for WGBS data and Methylation450K data available in \citep{ziller2013charting} and  \citep{zhang2015predicting}, respectively. We focus on the following criteria:
\begin{eqnarray*}
RMSE&=&\sqrt{\frac{\sum^{k^*}_{i=1}\sum^{n^*}_{j=1}(\hat y^*_i(s^*_j )-  y^*_i(s^*_j  ))^2 }{k^*n^*}},\, \\
{P_{CI}(95\%)} &=& \frac{1}{k^*{n^{*}}}\sum\limits_{i = 1}^{k^*} {\sum\limits_{j = 1}^{n^{*}} 1\{y^*_{i}( s^{*}_j)\in C{I_{ij}}(95\% )\}}\,, \\
{L_{CI}(95\%)} &=& \frac{1}{{k^*{n^{*}}}}\sum\limits_{i = 1}^{k^*} \sum\limits_{j = 1}^{{n^{*}}} {\Length\{C{I_{ij}}(95\% )\} } \,,
\end{eqnarray*}
where  for $1\leq i\leq k^*$ and $1\leq j\leq n^*$,  $y^*_i(s^*_j)$ is the held-out methylation levels of the $i^{th}$  sample  at the $j^{th}$ CpG site; $\hat y^*_{i}( s^{*}_j)$ is  the predicted  held-out methylation level of the $i^{th}$  sample  at the $j^{th}$ CpG site; $C{I_{ij}}(95\% )$ is the $95\%$ posterior credible interval; and $\Length\{C{I_{ij}}(95\% )\}$ is the length of the $95\%$ posterior credible interval.   An effective method is  expected  to have small out-of-sample RMSE, ${P_{CI}(95\%)}$ being close to nominal $95\%$ level, and  small $L_{CI}(95\%)$.  In \cite{zhang2015predicting}, a CpG site is defined to be methylated if  more than $50\%$ of the probes are methylated, and the accuracy of a method is defined by the proportion of the correct predictions of CpG sites being methylated or not.  We also include the rate  of accuracy as a criterion for comparison.

\vspace{-.1in}
\subsection{Application 1: WGBS data}
\label{subsec:WGBS}
 We first compare the out-of-sample prediction of different methods using the criteria discussed above for $10^6$ methylation levels at chromosome 1 in the WGBS  dataset \citep{ziller2013charting}. In this dataset,  24 samples are available in total and we randomly sample  $k^*=4$ samples,  whose  methylation levels are partially observed (with certain proportion being held out), while the methylation levels of the rest of the samples are fully observed.  We consider three scenarios, in which $25\%$, $50\%$, $75\%$, and  $90\%$ of the methylation levels of these $4$ samples are held out as the test dataset.   The methylation levels of each sample are centered and the mean is added  back for prediction in the GaSP model \citep{higdon2008computer}. 
 We estimate the range and nugget parameters  using equation (\ref{equ:max_marginal_post})  
and rely on the predictive distribution of model (\ref{equ:nonseparable1}) for combining different sources of information in prediction.

\begin{table}[t!]
\begin{center}
\begin{tabular}{lrrrr}
  \hline
        $25\%$ held-out  CpG sites                 & RMSE &$P_{CI}(95\%)$ & $L_{CI}(95\%)$ & Accuracy   \\
  \hline
  Nonseparable GaSP            &{.0835}&{.941}&{.258}&{.972} \\
  Nearest neighbor                   & .152 &/&/& .942 \\
  Linear  model by site  & .0993&.912&.261&.966  \\
  Random forest by site & .103&/&/& .964 \\
  Linear  model by sample & .100&.941&.308&.962   \\
  Random forest by sample  & .0961&/&/& .965 \\
  \hline
      $50\%$ held-out  CpG sites                 & RMSE &$P_{CI}(95\%)$ & $L_{CI}(95\%)$ & Accuracy   \\
  \hline
  Nonseparable GaSP            &{.0840}&{.943}&{.264}&{.971} \\
  Nearest neighbor                   & .153 &/&/& .942 \\
  Linear  model by site  & .0993&.913&.262&.966  \\
  Random forest by site & .103&/&/& .964 \\
  Linear  model by sample & .100&.941&.307&.962  \\
  Random forest by sample  & .0965&/&/& .964 \\
  \hline
      $75\%$ held-out CpG sites                  & RMSE &$P_{CI}(95\%)$ & $L_{CI}(95\%)$ & Accuracy   \\
  \hline
  Nonseparable GaSP            &{.0887}&{.939}&{.271}&{.969} \\
  Nearest neighbor                   & .166&/&/&.934  \\
  Linear  model by site  & .106&.910&.274& .963 \\
  Random forest by site & .108&/&/& .962 \\
  Linear  model by sample &.100  &.940&.308&.962  \\
  Random forest by sample  &.0972 &/&/&.964  \\
  \hline
      $90\%$ held-out CpG sites                  & RMSE &$P_{CI}(95\%)$ & $L_{CI}(95\%)$ & Accuracy   \\
  \hline
  Nonseparable GaSP            &{.0984}&{.917}&{.267}&{.965} \\
  Nearest neighbor                   & .188&/&/&.922  \\
  Linear  model by site  & .114&.908&.289& .959 \\
  Random forest by site &.114 &/&/&.958  \\
  Linear  model by sample& .100 &.941&.308&.962  \\
  Random forest by sample  &.0983 &/&/&.963  \\
  \hline

   \hline

\end{tabular}
\end{center}
   \caption{Comparison of different methods for $10^6$ methylation levels at chromosome 1 in the WGBS data. From the upper to the lower,  $25\%$, $50\%$, $75\%$ and $90\%$ of the first million methylation levels of $k^*=4$ samples are held out for testing, respectively.    }
   \label{tab:RMSE_Accuracy}
\end{table}

 As shown in Table~\ref{tab:RMSE_Accuracy}, the  nonseparable GaSP model has the smallest out-of-sample RMSE in out-of-sample prediction in almost all scenarios. For instance,  when $50\%$ of CpG sites are held out,   the nonseparable GaSP method improves the RMSE by around $15\%$  compared to any other methods we considered. The gain is from  integrating  the correlations  between CpG sites and between samples through a coherent statistical model, while the other models only utilize  partial    information. For example, the methods in rows 2 to 4 only exploit the site-wise correlation  by assuming the independence across different  samples  at each CpG site, while the methods in rows 5 to 6 only exploit the correlation between samples.

 When  more and more data  are held-out  as the test data,  the correlation between nearby observed methylation levels gets smaller as the average distance between two  sites with observed methylation levels  gets larger,  making it harder for prediction. We noticed the RMSE by the nonseparable GaSP model increases by more than $10\% $ when the percentage of held-out site increases from $75\%$ to $90\%$. In this scenario, the nonseparable GaSP model performs similar to the methods based on the sample correlation, and much better than the recent interpolation method based on the correlation between site \citep{zhang2015predicting}.

The nonseparable GaSP also leads to around $97\%$ accuracy in predicting whether a CpG site is methylated or not,  which is also the highest compared to all other methods.   Note the differences between the nonseparable model and other methods are small, as around $90\%$ of the CpG sites are methylated in this dataset, and thus  a benchmark estimator could achieve at least $90\%$ accuracy in prediction.

 \subsection{Application 2: Methylation450K data}
 \label{subsec:methy450}
\begin{table}[t]
\begin{center}
\begin{tabular}{lrrrr}
  \hline
                                                & RMSE &$P_{CI}(95\%)$ & $L_{CI}(95\%)$ & Accuracy   \\
  \hline
  Nonseparable GaSP            &{.0296}&{.958}&{.103} &{.991} \\
  Nearest neighbor          &.350 &/&/& .774 \\
  Linear model by site  &.0342 &.944& .099&.990  \\
  Random forest by site &.0339 &/&/&  .989\\
  Linear model by sample &.0304 &.957& .106&.990\\
  Random forest by sample  &.0304 &/&/& .989 \\
  \hline

\end{tabular}
\end{center}
   \caption{Predictive performance of different approaches for   the Methylation450K data.  }

   \label{tab:RMSE_Accuracy2}
\end{table}

In this section, we study the numerical performance of all the methods for all chromosomes the Methylation450K dataset \citep{zhang2015predicting}. In this dataset, the methylation levels at all chromosomes of $100$ samples are recorded, which contains $2\%$ methylation levels  in the WGBS whole sequencing dataset. Our ultimate goal is to interpolate the methylation levels at all chromosomes based on the Methylation450K dataset, so this dataset may give us a better sense of the performance of different approaches, compared to interpolation results at  chromosome 1 using the WGBS whole sequencing dataset shown in Section \ref{subsec:WGBS}.

The methylation levels at $n^*=7.4 \times 10^4$ CpG sites from $k^*=50$ samples are held out as the testing output, which is around $20\%$ of the total CpG sites in the Methylation450K dataset. We do not hold out more CpG sites because the observations in the Methylation450K dataset is already very sparse. In Table~\ref{tab:RMSE_Accuracy2}, the prediction based on the nonseparable GaSP model has the lowest RMSE, and around $95\%$ of the held-out data are covered in the $95\%$ predictive interval. Furthermore, around half of the probes are methylated in this dataset on average, so the predictive accuracy of the nonseparable GaSP is high compared with a benchmark estimator. 

Note that the CpG sites are sparse in the Methylation450K dataset, so the correlation of the methylation levels between nearby sites is smaller than the ones in the WGBS data. However, the long-ranged site-wise correlation is still useful for prediction. We found that the nearest neighbor method is the worst method among all, as it only employs the correlation at the nearest neighbor. The linear model by site and random forest by site employ the correlation on more neighboring sites, which improve the predictive accuracy. The nonseparable GaSP models both the correlation of the methylation levels between all sites and between all samples, leading to the most accurate predictions among the methods we compared.

The computation of the nonseparable GaSP model relies heavily on the fast and exact computation algorithm discussed in Section~\ref{sec:computation}.  Since the number of methylation levels is at the size of a million in one chromosome in one sample, direct computation of the GaSP model is  infeasible. In the supplemental materials, we also compare with a localized Kriging method. The performance of the nonseparable GaSP model is still better, partly because the correlation between CpG sites is long-ranged, shown in the left panel of Figure~\ref{fig:empirical_corr}.

\section{Concluding remarks}
\label{sec:conc}

This paper discusses modeling multiple functional data  through Gaussian stochastic processes. We unify several different models, including the linear regression model and separable model,  through a nonseparable GaSP framework.    A computationally efficient algorithm is provided for the large scale problems without approximation to the likelihood. Several interesting future topics are  worth exploring  from both the computational and modeling perspectives. The achievement in computation is limited, in a sense that the input of the GaSP model (CpG site) is only 1 dimensional. { It remains to be an issue to generalize this computational method for the case with multi-dimensional inputs. Some recent progresses of this direction are introduced in \cite{lindgren2011explicit}, where the GaSP with a Mat{\'e}rn covariance can be represented by stochastic partial differential equations, while a method that computes the exact likelihood is still unknown.}  
 Furthermore, the outcomes of methylation levels  are $[0,1]$ with lots of $0$ and $1$ in the dataset. One may model a point process with probability masses at 0 and 1 to further improve the accuracy in prediction.

 In this work, we estimate the factor loadings by the principal component analysis. Another data-dependent way to estimate the factor loadings is by the generalized probabilistic principal component analysis \citep{gu2018generalized}, where the factor processes are first marginalized out and the factor loadings are estimated by the maximum marginal likelihood estimator. Various studies considered the space-varying factor loadings through certain basis functions, such as local bisquare functions \citep{cressie2008fixed,ma2017fused}, wavelets \citep{shi2007global}, and orthonormalized cubic B-splines \citep{chu2014semiparametric}. It is interesting to compare the performance of these approaches. Furthermore, in estimating the factor loadings, we only use the observations on $\mathbf s^{\mathscr D}$. It may be more satisfying to develop a method to use all data, including $\mathbf y(\mathbf s^*)$, to estimate the factor loadings.

  \section*{Acknowledgements}  The research of Mengyang Gu was part of his PhD thesis at Duke University. The authors thank the editor, the associate editor and two
referees for their comments that substantially improved the article. The authors sincerely thank Barbara Engelhardt for providing the methylation level data and discussion.

\section*{Appendix: Proofs for Section~\ref{sec:models}}
\label{sec:proofs}
\begin{proof}[Proof of Lemma~\ref{lemma:lik_nonseparable1}]

  Because $\mathbf { A}=\mathbf U \mathbf D/\sqrt{n}$, $\mathbf A (\mathbf A^T \mathbf A)^{-1} \mathbf A^{T} =\mathbf I_{K}$. The likelihood  is

\begin{align*}
p(\mathbf  Y_{v}(\mathbf s^{\mathscr D} )| \mathbf v_v(\mathbf s^{\mathscr D}), \sigma^2_0 ) =&  (2\pi \sigma^2_0)^{-nK/2} \exp\left(-\frac{\left(\mathbf  Y_{v}(\mathbf s^{\mathscr D})-\mathbf A_{v} \mathbf v_v(\mathbf s^{\mathscr D}) \right)^T\left(\mathbf  Y_{v}(\mathbf s^{\mathscr D})-\mathbf A_{v} \mathbf v_v(\mathbf s^{\mathscr D}) \right) } {2\sigma_0^2}\right) \\
=&(2\pi \sigma^2_0)^{-nK/2}   \exp\left(-\frac{\left(  \hat {\mathbf v}_v(\mathbf s^{\mathscr D})-\mathbf v_v(\mathbf s^{\mathscr D})   \right )^T\mathbf A^T_v \mathbf A_v\left(\hat {\mathbf v}_v(\mathbf s^{\mathscr D})-\mathbf v_v(\mathbf s^{\mathscr D}) \right) } {2\sigma_0^2}\right),
\end{align*}
where $\hat {\mathbf v}_v(\mathbf s^{\mathscr D})=  (\mathbf A_v^T \mathbf A_v)^{-1}\mathbf A_v^T \mathbf Y_v(\mathbf s^{\mathscr D})  $.  $\mathbf A_v^T \mathbf A_v$ is a diagonal matrix because  
\[\mathbf A_v^T \mathbf A_v = \left( {\begin{array}{*{20}{c}}
   {(\mathbf I_{n} \otimes \mathbf a_1)^T(\mathbf I_{n} \otimes \mathbf a_1) } & {(\mathbf I_{n} \otimes \mathbf a_1)^T(\mathbf I_{n} \otimes \mathbf a_2)} & {...} & {(\mathbf I_{n} \otimes \mathbf a_1)^T(\mathbf I_{n} \otimes \mathbf a_{K})}  \\
   {(\mathbf I_{n} \otimes \mathbf a_2)^T(\mathbf I_{n} \otimes \mathbf a_1) } & {(\mathbf I_{n} \otimes \mathbf a_2)^T(\mathbf I_{n} \otimes \mathbf a_2)} & {...} & {(\mathbf I_{n} \otimes \mathbf a_2)^T(\mathbf I_{n} \otimes \mathbf a_{K})}  \\
   {...} & {...} & {...} & {...}  \\
   {(\mathbf I_{n} \otimes \mathbf a_{K})^T(\mathbf I_{n} \otimes \mathbf a_1) } & {(\mathbf I_{n} \otimes \mathbf a_{K})^T(\mathbf I_{n} \otimes \mathbf a_2)} & {...} & {(\mathbf I_{n} \otimes \mathbf a_{K})^T(\mathbf I_{n} \otimes \mathbf a_{K})}  \\
 \end{array} } \right),\]
with $(\mathbf I_{n} \otimes \mathbf a_i)^T(\mathbf I_{n} \otimes \mathbf a_i)=(\mathbf I_{n}^T \otimes \mathbf a_i^T) (\mathbf I_{n} \otimes \mathbf a_i)=(\mathbf I_{n}^T \mathbf I_{n})\otimes (\mathbf a_i^T \mathbf a_i)  $
  and $(\mathbf I_{n} \otimes \mathbf a_i)^T(\mathbf I_{n} \otimes \mathbf a_j)=(\mathbf I_{n}^T \mathbf I_{n})\otimes ( \mathbf a_i^T  \mathbf a_j)=\mathbf O $, where $\mathbf O$ is a matrix with each element being $0$.   Marginalizing out $\mathbf v_v(\mathbf s^{\mathscr D})$, one has 
 \begin{align*}
&p(\mathbf  Y_{v}(\mathbf s^{\mathscr D} )|\bm \sigma^2_{0}, \bm \tau^2_{1:K}, \bm \gamma_{1:K}, \bm \eta_{1:K} ) \\
=&\int p(\mathbf  Y_{v}(\mathbf s^{\mathscr D} )| \hat{\mathbf v}_v(\mathbf s^{\mathscr D}), \sigma^2_0 ) p( \hat{\mathbf v}_v(\mathbf s^{\mathscr D}) | \bm \tau^2_{1:K}, \bm \gamma_{1:K}, \bm \eta_{1:K} ) d\hat{\mathbf v}_v(\mathbf s^{\mathscr D})\\
=&|\mathbf A_v^T \mathbf A_v|^{-1/2}(2\pi)^{-nK/2}  | \bm \Sigma_v+\sigma^2_0( \mathbf A_v^T \mathbf A_v)^{-1} |^{-1/2}{\exp\left(-\frac{1}{2} {\mathbf {\hat v}_v(\mathbf s^{\mathscr D})}^T\left(\bm \Sigma_v+\sigma^2_0( \mathbf A_v^T \mathbf A_v)^{-1} \right)^{-1}{\mathbf {\hat v}_v(\mathbf s^{\mathscr D})}  \right) }\\
=&|\mathbf A_v^T \mathbf A_v|^{-1/2} \times \\
&\prod_{i=1}^{K} \left\{ (2\pi)^{-n/2} | \sigma^2_i \tilde{\mathbf R}_i +\sigma^2_0 (\mathbf a^T_i \mathbf a_i )^{-1}\mathbf I_{n} |^{-1/2}{\exp\left(-\frac{1}{2} {\mathbf {\hat v}_i(\mathbf s^{\mathscr D})}^T\left( \sigma^2_i \tilde{\mathbf R}_i +\sigma^2_0 (\mathbf a^T_i \mathbf a_i )^{-1}\mathbf I_{n}   \right)^{-1}{\mathbf {\hat v}_i(\mathbf s^{\mathscr D})}  \right) }\right\}.
\end{align*}
The last row follows from the fact that $\mathbf {\hat v}_i(\mathbf s^{\mathscr D})^T$ is the $l^{th}$ row of the matrix $\hat {\mathbf v}(\mathbf s^{\mathscr D}) $.
 \end{proof}

\begin{proof}[Proof of Lemma~\ref{lemma:pred_nonseparable1}]
Denote $\mathbf Y(\mathbf s^{\mathscr D}; s^*_j):=[\mathbf Y(\mathbf s^{\mathscr D});  Y(s^*_j) ] $ and  $\mathbf v(\mathbf s^{\mathscr D}; s^*_j):=[\mathbf v(\mathbf s^{\mathscr D});  v(s^*_j) ] $.   Both are $k \times (n+1)$ matrices. 

 Vectorizing the output $\mathbf  Y_{v}(\mathbf s^{\mathscr D}, s^*_j):=vec(\mathbf Y(\mathbf s^{\mathscr D}, s^*_j))$, a $K\times (n+1)$ vector, and $\mathbf v_v(\mathbf s^{\mathscr D}):=vec(\mathbf v(\mathbf s^{\mathscr D})^T )$, we can write model (\ref{equ:nonseparable1}) as, 
 \[ \mathbf  Y_{v}(\mathbf s^{\mathscr D}, s^*_j)= \mathbf A_{v} \mathbf v_v(\mathbf s^{\mathscr D},s^*_j)+\bm \epsilon, \]
  where $\bm \epsilon\sim \MN(\mathbf 0, \sigma_0^2 \mathbf I_{(n+1)K})$. Similar to the proof of Lemma~\ref{lemma:lik_nonseparable1}, one has 
\[ p( \mathbf  Y_{v}(\mathbf s^{\mathscr D}, s^*_j)\mid \bm \sigma^2_{0}, \bm \tau^2_{1:K}, \bm \gamma_{1:K}, \bm \eta_{1:K} )=|\mathbf A_v^T \mathbf A_v|^{-1/2}  \prod^{K}_{i=1} p_{ {MN}}(\hat {\mathbf  v}_i(\mathbf s^{\mathscr D}, s^*_j); \mathbf 0, \sigma^2_i \bm  \Lambda_i+\sigma^2_0 (\mathbf a^T_i \mathbf a_i)^{-1} \mathbf I_{n+1}) ), \]
where $\hat {\mathbf v}_i(\mathbf s^{\mathscr D}, s^*_j)$ is the transpose of the $l^{th}$ row of the $\hat {\mathbf v}(\mathbf s^{\mathscr D}, s^*_j): = (\mathbf A^T \mathbf A)^{-1}\mathbf A^T \mathbf Y(\mathbf s^{\mathscr D},s^*_j )$ and 
$\bm \Lambda_i= \left( {\begin{array}{*{20}{c}}
\tilde{\mathbf R}_i &  \mathbf r_i(s^*_j)\\
 \mathbf r_i^T(s^*_j)&  \tilde c_i(s^*_j,s^*_j)  \\
 \end{array} } \right).$ One has
 \begin{equation} 
 (\hat v_i(s^*_j)| \hat {\mathbf v}_i(\mathbf s^{\mathscr D}), \bm \sigma^2_{0},  \tau^2_i,  \gamma_i,  \eta_i) \sim  \N( \hat v^*_i(s^*_j), \sigma^2_i {\tilde c}^*(s^*_j)+\sigma^2_0 (\mathbf a^T_i \mathbf a_i)^{-1} ), 
 \label{equ:pred_v_sj}
 \end{equation}
 with $\hat v^*_i(s^*_j)=\mathbf r_i^T(s^*_j) \left( \tilde{\mathbf R}_i+ \frac{\sigma^2_0 (\mathbf a^T_i \mathbf a_i)^{-1}}{ \tau^2_i} \mathbf I_n\right)^{-1} \hat {\mathbf v}_i(\mathbf s^{\mathscr D})$ and $\tilde c_i^*(s^*_j)=\tilde c_i(s^*_j,s^*_j)-\mathbf r_i^T(s^*_j) (\tilde{\mathbf R}_i+ \frac{\sigma^2_0 (\mathbf a^T_i \mathbf a_i)^{-1}}{ \tau^2_i} \mathbf I_n)^{-1} \mathbf r_i(s^*_j)$. Note 
 $\mathbf A (\mathbf A^T \mathbf A)^{-1} \mathbf A^{T} =\mathbf I_{K}$. One has $\mathbf Y(s^*_j)=\mathbf A \mathbf {\hat v}(s^*_j)$ and $\mathbf Y(s^{\mathscr D})= \mathbf A\mathbf {\hat v}(s^{\mathscr D})$. Applying the properties of multivariate normal distribution to (\ref{equ:pred_v_sj}) leads to the results.

\end{proof}

  \bibliographystyle{apalike}
\bibliography{References_2018}

\newpage

\begin{center}
    {\bf \LARGE Supplementary materials}
\end{center}

\beginsupplement

\quad 

All the formulas in this supplementary materials are cross-referenced in the main body of the article.  

  \section{Closed form quantities of the continuous state space model}
\label{sec:closed_form_SDE}

We  give the quantities of continuous state space model representation in (\ref{equ:ctdlm}) in the main body of the article.

For $1\leq i\leq K$,  the SDE is \citep{hartikainen2010kalman}.
 \begin{align*}
  \frac{d\bm {\theta}_i(s)}{ds}=\mathbf J_i\bm {\theta}_i (s)+\sqrt{q_i}\mathbf L z_i(s),  
  \end{align*}
where $z_i(s)$ is a standard Gaussian white noise process, $q_i=16\sigma^2_i\lambda^5_i/3$,  $\mathbf L=(0,0,1)^T $, and 
 \[\mathbf J_i= \begin{pmatrix}
 0&1  &0 \\ 
 0&0  &1 \\ 
 -\lambda^3_i& -\lambda^2_i  &-3\lambda_i
\end{pmatrix}.  \]
  
 Denote $ d_j=|s_{j+1}-s_{j}|$ for $j=1,...,N-1$.  We have the following expressions for the solution of the SDE  in (\ref{equ:ctdlm}) in the main body of the article,
\[\mathbf G_i(s_j)= e^{\mathbf J_i  d_j}=\frac{e^{-\lambda_i  d_j}}{2} \begin{pmatrix}
 \lambda_i^2  d_j^2+2\lambda_id_j+2&2(\lambda_i  d_j^2+ d_j)  & d_j^2\\ 
 -\lambda_i^3 d_j^2&-2(\lambda_i^2  d_j^2-\lambda_i  d_j-1)  &2d_j-\lambda_i  d_j^2 \\ 
\lambda_i^4 d_j^2-2\lambda_i^3 d_j& 2(\lambda_i^3 d_j^2-3\lambda_i^2 d_j) &\lambda_i^2 d_j^2-4\lambda_i  d_j +2
\end{pmatrix} \]
\[\mathbf W_i(s_j) =\frac{4\sigma^2_i\lambda_i^5}{3}\begin{pmatrix}
W_{1,i}(s_j)  &W_{2,i}(s_j)   &W_{3,i}(s_j)   \\ 
W_{4,i}(s_j)  &W_{5,i}(s_j)  &W_{6,i}(s_j)   \\ 
W_{7,i}(s_j)  &W_{8,i}(s_j)  &W_{9,i}(s_j)
\end{pmatrix},  \]
with 
\begin{align*}
W_{1,i}(s_j)&=\frac{e^{-2\lambda_i  d_j} (3+6\lambda_i d_j+6\lambda_i^2 d^2_j+4\lambda_i^3 d^3_j+2\lambda_i^4 d^4_j)-3 }{-4\lambda_i^5}, \\
W_{2,i}(s_j)&=W_{4,i}(s_j)=\frac{e^{-2\lambda_i  d_j}d_j^4}{2}, \\
W_{3,i}(s_j)&=W_{7,i}(s_j)=\frac{e^{-2\lambda_i  d_j}(1+2\lambda_i  d_j +2\lambda_i^2  d^2_j+4\lambda_i^3 d^3_j-2\lambda_i^4 d^4_j )-1   }{4\lambda_i^3}, \\
W_{5,i}(s_j)&=\frac{e^{-2\lambda_i  d_j}(1+2\lambda_i  d_j +2\lambda_i^2  d^2_j-4\lambda_i^3 d^3_j+2\lambda_i^4 d^4_j )-1   }{-4\lambda_i^3}, \\
W_{6,i}(s_j)&= W_{8,i}(s_j)=\frac{e^{-2\lambda_i  d_j} d_j^2(4-4\lambda_i d_j+\lambda_i^2 d^2_j) }{2}, \\
W_{9,i}(s_j)&= \frac{e^{-2\lambda_i  d_j}(-3+10\lambda_i d_j-22\lambda_i^2 d^2_j+12\lambda_i^3 d^3_j-2\lambda_i^4 d^4_j)+3  }{4\lambda_i},
\end{align*}
and 
\[\mathbf W_i(s_{1})= \begin{pmatrix}
 \sigma^2_i&0  &-\sigma^2_i\lambda^2_i/3 \\ 
 0&\sigma^2_i\lambda^2_i/3  &0 \\ 
 -\sigma^2_i\lambda^2_i/3 &0 &\sigma^2_i \lambda^4_i 
\end{pmatrix}, \]

for $j=1,...,N$ and $i=1,...,K$.

For $i=1,...,K$, the joint distribution of $\bm \theta_i(s_{1:N})$ is given below 
\[
 \begin{pmatrix}
\bm \theta_i(s_1)  \\ 
\bm \theta_i(s_2)  \\ 
...\\
\bm \theta_i(s_N)  \\
\end{pmatrix} \sim  \MN\left( \mathbf 0_{3N}, \bm \Lambda^{-1} \right), \]
where 
\begin{align*}
\setlength\arraycolsep{2.1pt}
\scriptsize
\bm \Lambda=\begin{pmatrix}
\mathbf W^{-1}_i(s_{1}) +\mathbf G^T_i(s_2)\mathbf W^{-1}_i(s_2)\mathbf G_i(s_2) &-\mathbf G_i^T(s_2)\mathbf W^{-1}_i(s_2)&& &\\ 
-\mathbf W^{-1}_i(s_2)\mathbf G_i(s_2) &\mathbf W^{-1}_i(s_{2}) +\mathbf G^T_i(s_3)\mathbf W^{-1}_i(s_3)\mathbf G_i(s_3) &&&  \\ 
&-\mathbf W^{-1}_i(s_3)\mathbf G_i(s_3)&...&& \\
&...&...&-\mathbf G_i^T(s_N)\mathbf W^{-1}_i(s_N)\\
 &...&-\mathbf W^{-1}_i(s_N)\mathbf G_i(s_N)&\mathbf W^{-1} _i(s_{N}) \\
\end{pmatrix}
 \end{align*}

  \section{Combing feature data into the nonseparable model}
  \label{sec:regressors}

 To impute the methylation levels,  some  site-specific features such as genomic position, DNA sequence properties, cis-regulatory element, can be used as covariates in a regression model.  Incorporating  regressors/covariates is less studied in the nonseparable GaSP model.   In this section, we  discuss  a way to  jointly model the site-specific  features and output.

 Let $\mathbf X(s)_{[q\times 1]}$ be  features at site $s$ (including the intercept).   Consider an extended model 
      \begin{equation}
    \mathbf Y^e(s)=\mathbf A^e  \mathbf {\tilde v}(s)+\bm \epsilon_0, 
        \label{equ:nonseparable2_e}
    \end{equation}
 for every $s \in \mathscr S$, where $ \mathbf Y^e(s)=(\mathbf X^T(s);\mathbf Y^T(s)  )^T$,  the weight $\mathbf {\tilde v}(\cdot)$ is  defined the same as  in  (\ref{equ:GaSP}) in the main body of the article,  and $\bm \epsilon_0 \sim {\MN}(0,\sigma^2_0\mathbf I_{K+q})$. Let $\mathbf X(\mathbf s^{\mathscr D})_{[q\times n]}$ be the features at sites $\mathbf s^{\mathscr D}$ and   $\mathbf Y^e(\mathbf s^{\mathscr D})= (\mathbf X^T(\mathbf s^{\mathscr D});\mathbf Y^T(\mathbf s^{\mathscr D})  )^T$. The extended basis matrix $\mathbf A^e=\mathbf U^{e}\mathbf D^{e}/\sqrt{n}$, where $\mathbf U^{e}$ and $\mathbf D^{e}$ are still defined through the SVD decomposition $\mathbf Y^e(\mathbf s^{\mathscr D})= \mathbf U^{e}\mathbf D^{e}\mathbf V^{e} $. The predictive distribution of the model (\ref{equ:nonseparable2_e}) can be obtained similarly to Lemma~\ref{lemma:pred_nonseparable1}.

  The connection among regression model, the separable GaSP model, and the nonseparable GaSP model shown in the previous section still holds.    
  Let $\mathbf Z^e$ follow a matrix normal distribution with mean zero and   covariance $\bm \Sigma^e \otimes \mathbf \Lambda$, where $\bm \Sigma^e= \left( {\begin{array}{*{20}{c}}
\bm \Sigma^e_{00} & \bm \Sigma^e_{0*}\\
 \bm \Sigma^e_{*0}& \bm  \Sigma^e_{**}  \\
 \end{array} } \right)$ is  a $(q+K)\times (q+K)$ covariance matrix.  The separable model is a special case of the nonseparable GaSP model (\ref{equ:nonseparable2_e}) specified in Remark~\ref{remark:nonseparable_separable}.  The prediction by the  regression model  (\ref{equ:regress_ppl}) with covariates $\mathbf H_i(s_j)=  ( \mathbf X^T(s_j);\mathbf y^T(s_j) )^T$ is also a special case of the extended nonseparable GaSP model specified in Remark~\ref{remark:nonseparable_LM}.

\section{Comparison to approximation method by blocks}
\label{sec:approximation}

In this subsection, we compare our exact and fast computation of the nonseparable GaSP model with a straightforward approximation, in which the long sequence is divided into small blocks and GaSP models are built independently in each block. Assume the data are divided into $M$ blocks, each with $n_0$ inputs (where $n=Mn_0$), the computational operations of which are then $O(M n_0^3)$ instead of $O(n^3)$ for the inversion of the covariance matrix.

For illustration purposes,  the data are divided into 100 batches and 200 CpG sites are used as the training data in each batch. We consider two scenarios, with 600 CpG sites and $66$ CpG sites between these 200 CpG sites being selected as the test CpG sites in each batch, respectively. That means that  roughly 75\% and 25\%  of the data are  held out.   We still assume the methylation levels for the first 20 samples are  available at all CpG sites and  4  randomly selected samples are only partially observed. The total number of the test CpG sites is $240,000$  and $26,400$,  respectively.  

\begin{table}[t]
\begin{center}
\begin{tabular}{lrrrr}
  \hline
     $25\%$ held-out  CpG sites        & RMSE &$P_{CI}(95\%)$ & $L_{CI}(95\%)$ & Accuracy   \\
  \hline
  Nonseparable GaSP full model         & {0.083} & {0.956}&{0.278} & {0.969} \\
  Nonseparable GaSP by batch         &0.091&0.923&0.258&  0.966 \\
    \hline
     $75\%$ held-out  CpG sites        & RMSE &$P_{CI}(95\%)$ & $L_{CI}(95\%)$ & Accuracy   \\
  \hline
  Nonseparable GaSP full model           &{0.087}&{0.963}&{0.305} &{0.968} \\
  Nonseparable GaSP by batch         &0.096 &0.934&0.283&0.966 \\
\hline
\end{tabular}
\end{center}
   \caption{Comparison of different methods in terms of out of sample prediction for WGBS data with $25\%$  and $75\%$ CpG sites are held out for testing.   }

   \label{tab:RMSE_by_batch}
\end{table}

\begin{figure}[t]
\centering

	\includegraphics[height=.4\textwidth,width=.8\textwidth]{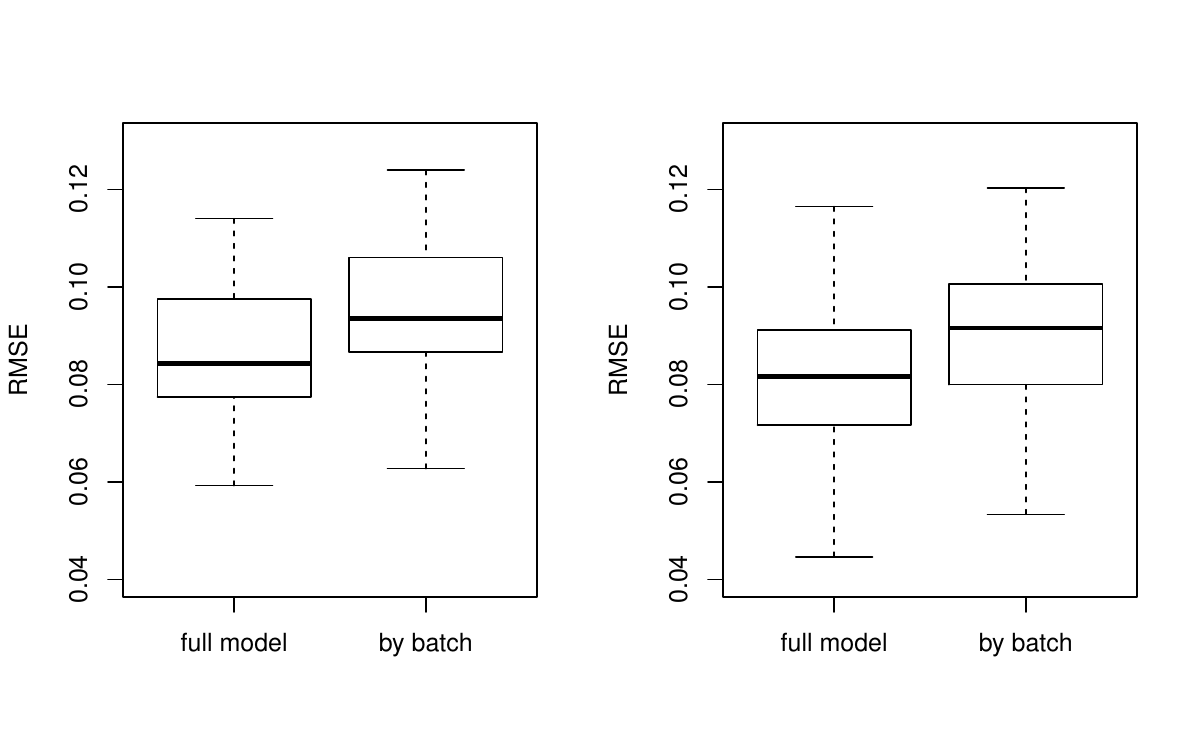}

   \caption{Boxplots of RMSE$_j$  for the  test samples in batch $j$  with $75\%$ of CpG sites (left panel) and $25\%$ of CpG sites (right panel) being held-out, respectively, $j=1, \dots, 100$. }
\label{fig:box_plot_by_batch}
\end{figure}

As shown in Table~\ref{tab:RMSE_by_batch}, the prediction by the nonseparable GaSP with the full model is about 10\% better in terms of RMSE in both scenarios. This is further justified by Figure~\ref{fig:box_plot_by_batch}.  One possible reason is that the boundary effect is large in the approximation method when we divide the data into batches.
 All these results suggest that simply approximating the likelihood by batch yields inferior predictive results than the nonseparable GaSP model with the full likelihood. Again, the implementation of the full model relies on the FFBS algorithm discussed in Section~\ref{sec:computation} in the main body of the article.


\end{document}